\documentclass[10pt,journal,compsoc]{IEEEtran}

\usepackage[ruled,vlined]{algorithm2e}
\usepackage{algorithm2e,setspace}
\usepackage{amsmath,algpseudocode}
\usepackage[para]{threeparttable}
\usepackage{amsmath,amssymb,amsfonts}
\usepackage{multirow}
\usepackage{graphicx}
\usepackage{textcomp}
\usepackage{xcolor}
\def\BibTeX{{\rm B\kern-.05em{\sc i\kern-.025em b}\kern-.08em
    T\kern-.1667em\lower.7ex\hbox{E}\kern-.125em}}

\usepackage{tcolorbox}
\usepackage[export]{adjustbox}
\usepackage{subcaption}
\usepackage{tikz}
\usepackage{pgfplots}
\usetikzlibrary{matrix}
\usepgfplotslibrary{groupplots}
\pgfplotsset{compat=1.12}
\usetikzlibrary{patterns}

\usepackage{mathtools}

\tikzset{
chart/.style={
  legend label/.style={font={\scriptsize},anchor=west,align=left},
  legend box/.style={rectangle, draw, minimum size=5pt},
  axis/.style={black,semithick,->},
  axis label/.style={anchor=east,font={\tiny}},
  },
pie chart/.style={
  chart,
  slice/.style={line cap=round, line join=round, very thick,draw=white},
  pie title/.style={font={\bfseries}},
  slice type/.style 2 args={
    ##1/.style={fill=##2},
    values of ##1/.style={}
    }
  }
}
\pgfdeclarelayer{background}
\pgfdeclarelayer{foreground}
\pgfsetlayers{background,main,foreground}

\newcommand{\pie}[3][]{
\begin{scope}[#1]
\pgfmathsetmacro{\curA}{90}
\pgfmathsetmacro{\r}{1}
\def\c{(0,0)}
\node[pie title] at (90:1.3) {#2};
\foreach \v/\s in{#3}{
    \pgfmathsetmacro{\deltaA}{\v/100*360}
    \pgfmathsetmacro{\nextA}{\curA + \deltaA}
    \pgfmathsetmacro{\midA}{(\curA+\nextA)/2}

    \path[slice,\s] \c
        -- +(\curA:\r)
        arc (\curA:\nextA:\r)
        -- cycle;
    \pgfmathsetmacro{\d}{max((\deltaA * -(.4/50) + 1) , .5)}

    \begin{pgfonlayer}{foreground}
    \path \c -- node[pos=\d,pie values,values of \s]{$\v\%$} +(\midA:\r);
    \end{pgfonlayer}

    \global\let\curA\nextA
}
\end{scope}
}

\ifCLASSOPTIONcompsoc
  \usepackage[nocompress]{cite}
\else
  \usepackage{cite}
\fi

\begin{document}

\title{SMURF: Efficient and Scalable Metadata Access for Distributed Applications}

\author{Bing~Zhang,
        Tevfik~Kosar,~\IEEEmembership{Senior~Member,~IEEE}
\IEEEcompsocitemizethanks{\IEEEcompsocthanksitem B. Zhang is with National Center for Supercomputing Applications University of Illinois at Urbana-Champaign, Champaign, IL 61801\protect\\
E-mail: bing@illinois.edu
\IEEEcompsocthanksitem T. Kosar is with the Department of Computer Science and Engineering at the University at Buffalo (SUNY), Buffalo, NY, 14260.\protect\\
}}


\IEEEtitleabstractindextext{%
\begin{abstract}
In parallel with big data processing and analysis dominating the usage of distributed and cloud infrastructures, the demand for distributed metadata access and transfer has increased. In many application domains, the volume of data generated exceeds petabytes, while the corresponding metadata amounts to terabytes or even more.
This paper proposes a novel solution for efficient and scalable metadata access for distributed applications across wide-area networks, dubbed SMURF. Our solution combines novel pipelining and concurrent transfer mechanisms with reliability, provides distributed continuum caching and prefetching strategies to sidestep fetching latency, and achieves scalable and high-performance metadata fetch/prefetch services in the cloud. 
We also study the phenomenon of semantic locality in real trace logs, which is not well utilized in metadata access prediction. We implement a novel prefetch predictor based on this observation and compare it with three existing state-of-the-art prefetch schemes on Yahoo! Hadoop audit traces.
By effectively caching and prefetching metadata based on the access patterns, our continuum caching and prefetching mechanism significantly improves local cache hit rate and reduces the average fetching latency. We replayed approximately 20 Million metadata access operations from real audit traces, in which our system achieved 90\% accuracy during prefetch prediction and reduced the average fetch latency by 50\% compared to the state-of-the-art mechanisms.
\end{abstract}

}

\maketitle

\IEEEdisplaynontitleabstractindextext
\IEEEpeerreviewmaketitle

\IEEEraisesectionheading{\section{Introduction}}

\label{sec:introduction}

\IEEEPARstart{W}{e} are witnessing a new era that offers new opportunities to conduct data-intensive scientific research with the help of recent advancements in computational, storage, and network technologies. With the rapid deployment of distributed infrastructures and the collaborations between different organizations (e.g., XSEDE~\cite{towns2014xsede}, OSG~\cite{pordes2007open}, Chameleon~\cite{Mambretti:2015:NGC:2919335.2920485} and Cloudlab~\cite{cloudlab2019}), it is feasible and promising to run scientific applications on these large-scale geo-distributed infrastructures. 
In many application domains, including environmental and coastal hazard prediction, climate modeling, high-energy physics, astronomy, and genome mapping, the volume of data generated has already exceeded petabytes, while the corresponding metadata amounts to terabytes or even more~\cite{weil2004dynamic}.
According to Roselli~\cite{roselli2000comparison}'s study, more than 50\% of all I/O operations are due to metadata-intensive computing, and the requests to read file attributes dominate in all workloads. 
The data movement is the common operation between the data I/O nodes, compute clusters, and user workstations for reconstruction, analysis, and visualization of the data. The cloud-hosted metadata catalog (e.g., Globus Catalog~\cite{wozniak2015big}, iRODS Metadata Catalog~\cite{schnase2017merra}) mitigates the difficulty of browsing, tracking, and discovery of the data. Thus, remote metadata retrieval and searching always have been conducted frequently between the users and cloud services, even over wide-area networks. 
Data lakes~\cite{hai2016constance, quix2016gemms, terrizzano2015data} have been proposed to meet the requirement that scientists and researchers are seeking broader access to different types of ``raw data'' organized in a contextual format that can be used across different projects. In contrast to the data warehouse schema~\cite{chaudhuri1997overview}, the data schema in a data lake is not predefined. With the help of a metadata description, a data lake system can annotate, integrate, and query the raw data. Without the metadata, data alone is not useful, and the data lake becomes a data swamp~\cite{hai2016constance}.

More recently, with the unprecedented growth of the Internet of Things (IoT) devices (e.g., sensors, virtual reality, smartphones, smart vehicles, smart homes, and smart grocery stores) connecting to the world ~\cite{ashton2009internet}, the drops in the cost of sensors~\cite{jankowski2014internet}, and the advanced technologies in wired and wireless networks, more than 50 billion edge devices are expected to be connected to the cloud by 2022~\cite{evans2011internet}. IoT devices autonomously capture and ingest data and seamlessly integrate with the modern IT infrastructures, and it is tenable to argue that IoT data is becoming the Big Data. In the IoT data processing, the real-time response is the inherent core feature. IBM estimates that 90\% of the data generated by the end devices like tablets and smartphones is never analyzed, and as much as 60\% of this data will start to lose value within milliseconds of being generated~\cite{IBM60}. Intelligent IoT applications such as camera-based monitoring systems collect the real-time data and send the aggregated information of the raw data to the remote analytic platforms (e.g., Apache Storm~\cite{Storm}, Apache Spark~\cite{zaharia2016apache}, and IBM Infosphere Streams~\cite{biem2010ibm}) for real-time decision-making. Traditional remote exchange of information from the distant cloud cannot meet the ultra-low latency requirements of these time-sensitive and geographically dispersed IoT applications. Consequentially, the challenge beckons a paradigm shift in which the data and metadata can be accessed anywhere, anytime, and from any device. 

Access to proper metadata is extremely latency-sensitive also due to user experience and critical business operations: Google reported 20\% revenue loss due to a specific experiment that increased the time to display search results by as little as 500 milliseconds; and Amazon reported 1\% sales decrease for an additional delay of as little as 100 milliseconds~\cite{Kohavi:2007:PGC:1281192.1281295}.
Unfortunately, most of the existing studies have focused on the efficient and scalable transfer of large-scale data, and there has been little work focusing on the optimization of remote access and transferring of metadata~\cite{zhang2015dls} in wide-area networks. By considering wide-area network latency, the frequency of revalidation of metadata, and the rapid growth of IoT, efficient and scalable metadata access and transfer technologies are demanded and expected to become a cornerstone of modern distributed IT infrastructures.


%

\begin{figure}[t]
  \centering
  \includegraphics[width=1\linewidth]{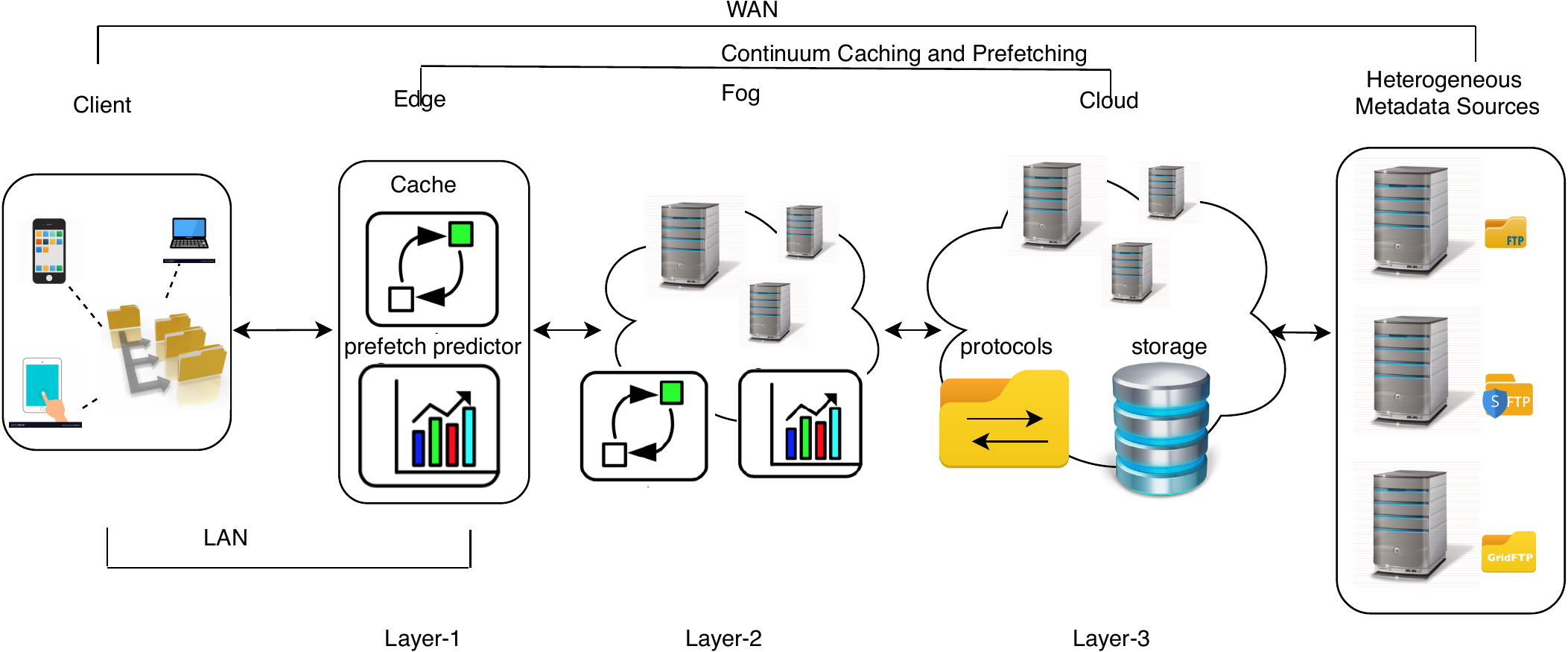}
  \caption{Client devices fetch/prefetch metadata of interest in WAN via SMURF's distributed continuum cache and prefetch mechanism.}
  \label {fig:usercase}
\vspace{-4mm}
\end{figure}

In this paper, we present a novel metadata access and retrieval system, called SMURF, which is built on the distributed continuum caching and prefetching architecture to effectively fetch, prefetch, and cache metadata on different hierarchical layers (as shown in Figure~\ref{fig:usercase}) between clients and remote I/O servers in a wide-area network (WAN) setting. The merits of the SMURF system have been illustrated by the scalability of the real-time metadata transferring and the reliability of metadata access from heterogeneous remote I/O nodes. The main contributions of this paper include:

\begin{itemize}

\item Design and implementation of an efficient and scalable metadata access and transferring technique for millions of metadata instances and records in WAN.

\item Design and implementation of a distributed continuum caching and prefetching technique to sidestep metadata access and transfer latency in WAN.

\item An interoperable solution that can work for heterogeneous metadata sources, not just specific for any protocol.

\item A study of the phenomenon of semantic locality in real system traces and development of a novel semantic locality prefetch scheme, which can achieve more than 90\% accurate prediction rate.

\item Comparison of four different prefetch predictors (our semantic locality prefetch predictor and three state-of-the-art predictors, namely, NEXUS, AMP and FARMER) and the legacy LRU cache on the Yahoo HDFS traces.


\end{itemize}

The rest of the paper is organized as follows: Section 2 describes the proposed system architecture and discusses its design issues; Section 3 presents the simulation methodology and performance evaluations; Section 4 discusses existing relevant work in this area; and Section 5 concludes the paper.

\section {System Architecture}

Two major goals of the SMURF system are (1) providing {\em interoperability} between heterogeneous and distributed nodes through on-the-fly inter-protocol translation and (2) improvement of the metadata transfer performance while meeting the {\em scalability} demands to enable large-scale metadata access over WAN on demand. Both of these capabilities are crucial in translating raw data into knowledge and discovery in an efficient way. 


Interoperability is a critical need since valuable data may reside on different I/O servers or cloud services. In such heterogeneous and distributed environments, users have to install different protocol clients (e.g., HTTP~\cite{rescorla2000http}, FTP~\cite{postel1985rfc}, GSIFTP~\cite{aloisio2002early}, IRODS~\cite{IRODS}, and Amazon S3~\cite{Palankar:2008:ASS:1383519.1383526}) which makes the edge devices tightly coupled with specific services. Therefore, it is cumbersome and requires extra expertise to switch between the different services. SMURF deploys a cloudlet edge cluster to the proximity of IoT devices, where the edge application can communicate with our edge cloudlet services with the universal programming interface (e.g., RESTful API), and the remote metadata resources to be retrieved can be expressed as Uniform Resource Identifier (URI) inside the requests.

Scalability is another central technical challenge in distributed metadata access. IoT applications, especially the sensor-based applications, have to process the dynamic workloads in real-time. The scalability of transferring large-scale metadata is a significant criterion to evaluate the performance of such systems. SMURF improves metadata transfer performance and meets scalability demands by using optimized pipelining and concurrency techniques.  Concurrency is an effective way to improve the end-to-end data transfer performance and has been well studied by many previous works~\cite{kosar2009new, yildirim2012end, liu2010data, deelman2006makes, kola2004disc, kosar2011stork, kim2015highly, yildirim2012gridftp}. Our system provides cloud service to establish and maintain multiple connections to the remote servers from a number of cluster machines. 
It also utilizes a hierarchical mechanism for the continuum caching and prefetching along the IoT-to-Cloud path, where the cache and the metadata prefetch predictor have been installed on each edge/fog layer.
SMURF employs a novel approach based on semantic locality to predict the metadata access of distributed workloads over WAN.

In the following subsections, we introduce the details of the SMURF architecture, discuss each functional component of the system, and outline the system's end-to-end operation workflow.

\subsection {SMURF Overview}
SMURF has a hierarchical architecture, as shown in Figure~\ref{fig:usercase}, consisting of two major components: (1) a centralized cloud cluster with the scalable fetch/prefetch services provides the universal pipelining and concurrent metadata transfer mechanisms with reliability; (2) the distributed continuum caching and smart prefetching strategies are deployed on edge/fog layers in clients' nearby networks, where the customized locality prefetch schemes utilize the local storage effectively to capture the future metadata to the proximity of clients.

\begin{figure}[t]
\begin{center}
\includegraphics[width=1\linewidth]{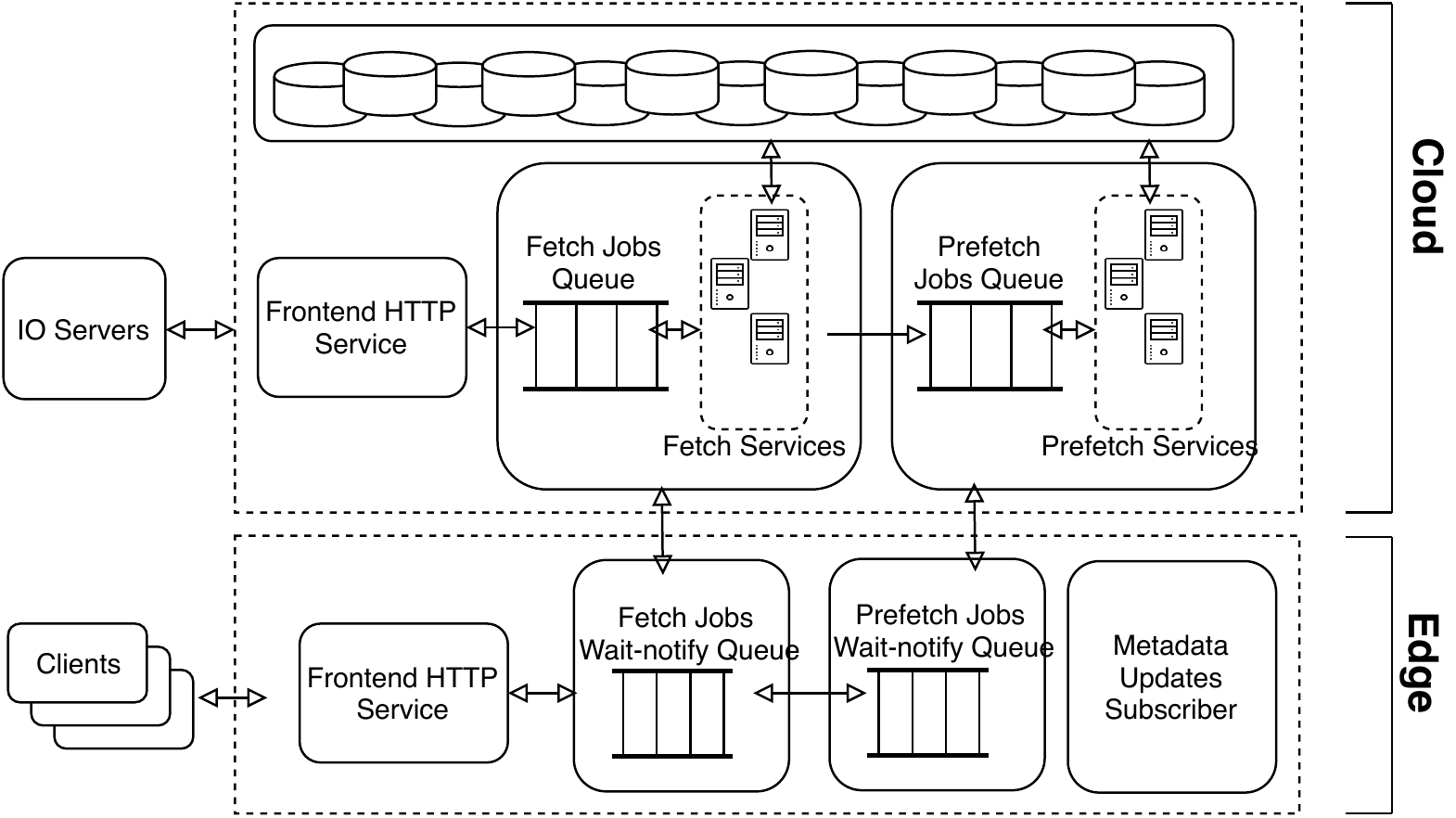}
\caption{ High-level overview of SMURF's metadata fetching and prefetching
between edge server and cloud.}
\label {overview-cloudedge}
\end{center}
\vspace{-4mm}
\end{figure}

Figure~\ref{overview-cloudedge} shows the high-level metadata fetching and prefetching mechanism between the edge and the cloud. When clients submit the metadata fetch requests to the edge server, the edge server will first try to read the metadata from the local storage and reply to the clients. The edge server will then send the requests to the prefetch predictor to analyze the request access patterns and predict the correlation metadata. The edge server establishes connections to the cloud server and sends/receives the fetch request and its correlation prefetch requests in pipelining. The edge requests will be dispatched to the clusters of fetch/prefetch services in the cloud and processed in parallel.


\subsection {Universal Metadata Transfer Stream}\label{transferstream}
The universal transfer stream is designed and implemented to coordinate and optimize the metadata access and transfer over different heterogeneous application-level transfer protocols over WAN.
One universal transfer stream retrieves the metadata of interest from the remote I/O server using a single TCP connection, and its novelty will be illustrated in three aspects. First, very different from implementing the traditional programming model, our implementation decouples the protocol's definition and the message transfer mechanism. We abstract and reconstruct the definition of protocol request as a chain of commands and parsers, and our transfer stream can send and parse different protocol requests in a universal mechanism. 
Furthermore, the protocol definition is provided as a library with programmability and extensibility; thus, users can follow the convention to customize their protocol to interact with the SMURF's universal transfer stream. Currently, SMURF supports application-layer transfer protocols such as FTP~\cite{postel1985rfc}, SFTP~\cite{moore1994transportation}, GSIFTP~\cite{allcock03}, IRODS~\cite{IRODS}, and Amazon S3~\cite{Palankar:2008:ASS:1383519.1383526}. Second, our transfer stream can efficiently utilize the network bandwidth via metadata transfer pipelining. The value of pipeline capacity can configure transfer stream channel, where the \textit{pipelining} capacity defines the maximum value of $C$ requests (request is a user's logical activity, such as auth, login, and metadata retrieval) to be continuously sent over one TCP connection without blocking or waiting for the completion of the previous replies from remote servers. Third, the stream is aware of the transfer status and supports failure recovery. When the connection is broken, the stream can automatically re-establish the connection and notify the service to re-dispatch the pending requests.

\subsubsection {Metadata Transfer Stream Programming Model}
An application-level protocol defines how the application exchanges the information between its distributed components. Especially during end-to-end metadata transferring, one round of information exchange will be initiated by a command sent from a client to the remote server, waiting for the arrival of the reply from the remote server, and then terminated by parsing the reply based on the protocol definition. The completion of one metadata request will require at least one round of message exchange between the client and the server. This whole process can be formally expressed in $Traditional(Request) = f_{1}(c_{1}) \circ f_{2}(c_{2}) \dotsm f_{n}(c_{n})$, where the completion of one request takes $n$ rounds of message exchanges and the notion of $f$ denotes a function to pack/send the protocol command $c$ and parse the reply. The operator $\circ$ concatenating two adjacent functions defines the order of message sending, receiving and parsing, thus it can denote the strict dependent relationship between two adjacent functions $f_{i} \circ f_{i+1} \implies  f_{i+1}  = g(f_{i} )$ where the sending of current message depends on the reply of previous messages. Moreover, the operator $\circ$ will take at least one round-trip-time (RTT) between two dependent adjacent command transmissions.

We can always decompose the function $f$ into two parts: message sending $s$ and message parsing $p$. Both functions of $s$ and $p$ can take a list of input parameters and 
apply the function execution over each element inside the given parameters. When the protocol definition does not require the dependency, then the pipeline transferring of one metadata request can be expressed as $Pipeline(request) = s(c_{1}, c_{2}, \dotsm , c_{n}) \circ p(c_{1}, c_{2}, \dotsm , c_{n})$. This expression follows three constraints: (1) The sender $s$ can continuously send request commands in the sequence order of $c_{1}, c_{2}, \dotsm, c_{n}$ without blocking, and meanwhile, the parser $p$ will parse the incoming replies in the same sequence order. 
(2) The operator $\circ$ strictly guarantees that the order of sending a command $c_{i}$ happens before parsing the reply of this command $c_{i}$, namely, $order(s(c_{ i})) < order(p(c_{i}))$. The overlapping executions of sending messages and parsing replies can be denoted as $s(c_{k} \dotsm c_{n}) \cap p(c_{1} \dotsm c_{k-1})$. It is possible because of the settings of the pipeline levels and the value of the RTT between the client and the server. (3) The completion of $Pipeline(request)$ will take at least one RTT.

If the protocol is stateless and requires the independent relationship between two adjacent commands' sending and receiving, the pipeline transferring of multiple $m$ requests over one stream will be expressed as (assuming transfer of the same type of requests, each of which consists of $n$ commands): 

\begin{equation*}
\begin{aligned}
Pipeline(R_{1}, R_{2}, \dotsm , R_{m}) = \\
s(c_{11}, c_{12}, \dotsm , c_{1n}) &\circ p(c_{11}, c_{12}, \dotsm , c_{1n}) \\
					     s(c_{21}, c_{22}, \dotsm , c_{2n}) &\circ p(c_{21}, c_{22}, \dotsm , c_{2n}) \\
					     &\vdotswithin{\circ} \notag \\
					     s(c_{m1}, c_{m2}, \dotsm , c_{mn}) &\circ p(c_{m1}, c_{m2}, \dotsm , c_{mn})
\end{aligned}
\end{equation*}

This expression resembles a matrix where the sender(s) and parser(s) of one request $R_{i}$ have been defined as row-wise, and the order of processing the requests has been defined as column-wise. The operator $ \circ$ still follows the aforementioned notion to define the order between the sender $s$ and the parser $p$ on the same row. Moreover, the overlapping executions of sending messages and parsing will happen across the requests, increasing the pipeline system throughput.

If the protocol is stateful that requires the dependent relationship between the commands' sending and parsing, then the transfer of one request cannot be interleaved by other requests. In this scenario, the pipeline transferring cannot give outstanding performance, but the system can still increase the transfer performance via concurrency. Namely, the system establishes multiple isolated connections to the remote server and transfers metadata messages simultaneously. Concurrent transferring will be discussed in more detail in section ~\ref{smurfcluster}.

To maximize the pipeline system's performance, the real-time stream transfer is preferable to the batch transfer. The new request should be put into the pipeline system immediately for transferring as long as the pipeline capacity is not full. Meanwhile, one request should be removed from the pipeline system once it is completed successfully or aborted. This real-time design of the pipeline system still needs to guarantee that the parser ordering is consistent with the sender ordering, which will be discussed in section ~\ref{matrixordering}.

\begin{algorithm}
\setstretch{1.10}
\caption{Send Metadata Requests}\label{algo:sendmetadatarequest}
\scriptsize

\begin{algorithmic}[1]
\State $channel$ $\leftarrow$ \Call{MetaChannel}{$host$, $port$, $pipelineCapacity$}
\State $protocolType$ $\leftarrow$ FTP, GSIFTP, IRODS ...
\State $channel$\Call{.open}{$protocolType$}

\item[]

\State $request$ $\leftarrow$ \Call{Request}{\null} 

\State $request$\Call{.authenticate}{$channel$, $credentials$}

\State $dependent$ $\leftarrow$ $True$ $or$ $False$

\State $request$\Call{.setDependentChain}{$dependent$}

\State $wait$ $\leftarrow$ $True$ $or$ $False$

\State $channel$\Call{.send}{$request$, $wait$}

\item[]

\State $request$ $\leftarrow$ \Call{Request}{\null}
\State $request$\Call{.list}{$path$}

\State $response$ $\leftarrow$ $channel$\Call{.send}{$request$, $wait$}

\item[]

\uIf{$response.wait = True$} {
    \State \Call{print}{$response$}
}
\end{algorithmic}
\end{algorithm}

\begin{algorithm}
\setstretch{1.10}
\caption{SMURF Protocol Request}\label{algo:protocolrequest}
\scriptsize

\begin{algorithmic}[1]
\Procedure{authenticate} {$channel$, $credentials$}
\State $cmdInfo$ $\gets$ \Call{packCred}{$credentials$}
\State $cmd$ $\gets$ \Call{Command}{$''auth''$, $cmdInfo$}
\State $parser$ $\gets$ \Call{ProtocolParser}{$request$}\Call{.authCmdParser}{$cmd$}

\State $pair$ $\gets$ \Call{Pair}{$cmd$, $parser$}
\State $pairs$\Call{.append}{$pair$}
\EndProcedure

\item[]

\Procedure{list} {$path$}
\State $cmdInfo$ $\gets$ \Call{packPath}{$path$}
\State $cmd$ $\gets$ \Call{Command}{$''list''$, $cmdInfo$}
\State $parser$ $\gets$ \Call{ProtocolParser}{$request$}\Call{.listCmdParser}{$cmd$}

\State $pair$ $\gets$ \Call{Pair}{$cmd$, $parser$}
\State $pairs$\Call{.append}{$pair$}
\EndProcedure

\end{algorithmic}
\end{algorithm}

\begin{algorithm}
\setstretch{1.15}
\caption{SMURF Protocol Parser}\label{algo:protocolparser}
\scriptsize

\begin{algorithmic}[1]
\Procedure{parse} {$reply$}
\State $myreply$ $\gets$ \Call{read}{$reply$}
\State $globalData1$ $\gets$ \Call{parseReply}{$myreply$}
\State $request$\Call{.save}{$globalData1$}
\State $globalData2$ $\gets$ $request$\Call{.get}{\null}

\State $cmdInfo$ $\gets$ \Call{packNext}{$reply$, $globalData2$}

\State $nextCmd$ $\gets$ \Call{Command}{$''next''$, $cmdInfo$}

\State $nextParser$ $\gets$ \Call{ProtocolParser}{$request$}\Call{.cmdParserNext}{$nextCmd$}

\State $request$\Call{.addPair}{$nextCmd$, $nextParser$}
\EndProcedure

\end{algorithmic}
\end{algorithm}


Algorithm~\ref{algo:sendmetadatarequest} shows the pseudocode for sending the metadata requests. First, the client establishes the metadata channel to the remote server (lines 1 - 3) with the provided information, such as host address, port number, and the pipeline capacity value. It also needs to provide the type of protocol to be used for metadata retrieval. Then an authentication request is generated with the given credentials (line 4). When sending the request, it is optional to specify that this client will do a blocking wait to complete authentication. Line 10 - 12 is to send the metadata request, namely, a listing request to retrieve the metadata content denoted by the resource path. Every metadata request can be sent in this convention, and the client can populate more requests into the metadata channel without waiting for the completion of the previous requests. Moreover, the metadata channel will automatically handle the commands' sending/parsing in the pipeline mechanism. 

In practice, any protocol can be written into SMURF protocol request and parser, as shown in Algorithms ~\ref{algo:protocolrequest} and ~\ref {algo:protocolparser}. In Algorithm~\ref{algo:protocolrequest}, the SMURF protocol request library packs the command message (line 3) and assigns the predefined parser to parse this reply (line 4). One request maintains a chain of pairs (line 6), where each pair is organized in the format of $\{command, parser\}$. The request decides the sequence order of commands in this pair chain and can append more pairs as the independent relationship (line 6). In Algorithm~\ref {algo:protocolparser}, each SMURF protocol parser can design and implement its logic to read the reply from the remote server (line 2). Parsers of this request share the data variables via the request space (lines 3 - 5). One parser can define the next dependent $\{command, parser\}$ based on the current result and append this pair into the pair chain (lines 6 - 9).


\subsubsection {Matrix Ordering Guarantees the Rule of ``You Parse What You Send''} \label{matrixordering}
\begin{figure}[t]
  \hspace{-3mm}
  \includegraphics[width=0.49\textwidth]{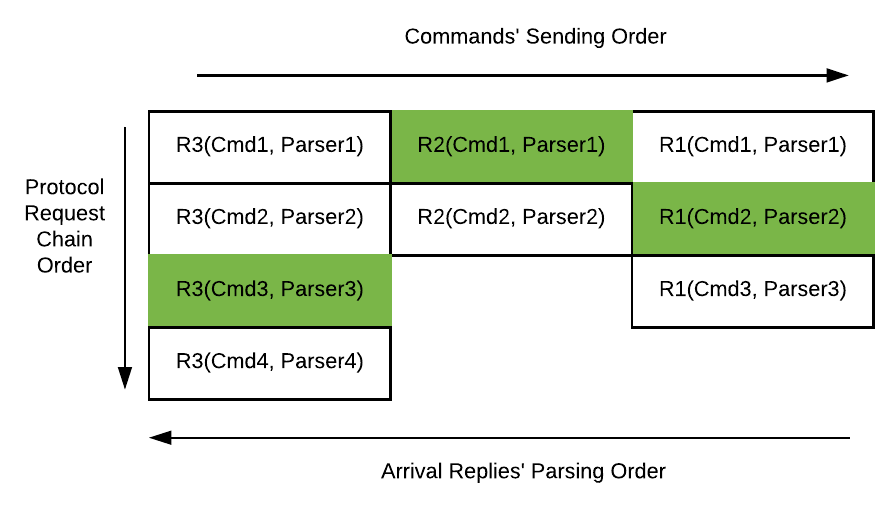}
  \caption{One possible scenario of matrix ordering to send/parse three requests in pipeline. The green color background denotes each request's inner cursor position to parse the arriving reply.}
  \label {fig:matrixordering}
\vspace{-4mm}
\end{figure}

Matrix ordering abstracts the messages' sending/parsing orders inside the universal transferring stream. In Figure~\ref{fig:matrixordering}, one request with the chain of pairs is expressed as a matrix column, where each pair $\{command, parser\}$ is a one-row element. Each row element also contains the information to specify whether this command will require the next row element's dependent relationship. 

The system can guarantee the correctness of the pipeline sending and parsing over multiple requests. Moreover, this correctness comes from two facts: (1) The underlying transfer connection guarantees that the order of sending messages is the same as the order of receiving the replies; (2) The matrix ordering guarantees that the order of parser in each pair is the same as sending its command message.

The system serves the message sending and parsing in parallel. The message sending is in Round Robin on the column order. We assume the request on the left-most column of the matrix will be served first without loss of generality. In Figure~\ref{fig:matrixordering}, followed by the sending of a command of $R_{1}$, then the commands of $R_{2}, R_{3}$ will be sent. When a new request comes, it will be placed on the left-most column, and the first command $Cmd_{1}$ from its chain will be sent immediately. Each request maintains an internal cursor pointing to the pair whose parser will be invoked for its incoming reply. When a request's parser has completed the arriving reply's execution, it will check the dependent relationship of the next pair on the below row. This parser will send the next command, and meanwhile, this request will be removed from the current column position and appended to the right-most. 
If there is no dependent relationship, this request will send all of its commands, and its parsers will continue to parse the arriving replies.  


The matrix ordering is thread-safe and can be interleaved by multiple threads. Figure~\ref{fig:matrixordering} demonstrates one possible scenario to process three stateless and dependent protocol requests: $R_{1}$ is a previous existing request, which finished its first command's sending and parsing and just sent its second command $Cmd_{2}$. $R_{1}$'s inner cursor is pointing to its $Parser_{2}$ and waiting for the reply. $R_{2}$ is a new request joining the system, and its first command has been sent, and thus its $Parser_{1}$ is waiting for the reply. $R_{3}$ is also a previous existing request, which already completed its first and second commands. Its third command has been sent, and thus $Parser_{3}$ is at the current inner cursor position and waiting for the reply. Overall, the system has sent the commands in the order of $\{$$R_{1}$$Cmd_{2}$, $R_{2}$$Cmd_{1}$, $R_{3}$$Cmd_{3}$$\}$ and will parse the arriving replies in the order of $\{$$R_{1}$$Parser_{2}$, $R_{2}$$Parser_{1}$, $R_{3}$$Parser_{3}$$\}$.

As shown in Algorithm 3, each parser has its own design and implementation to parse the incoming reply and decide the completion of parsing. Usually, this can be implemented as a state machine, where the transition of a state will be defined by the protocol's Request for Comments (RFC). For example, on retrieving the metadata of a filesystem folder containing millions of subfiles from a GSIFTP server, SMURF's transferring stream will receive a continuous intermediate part of metadata. The whole transferring will be terminated successfully when the parser can parse {\em code 250} of the reply. When one request's last parser has been completed, and then this request is finished. One request will be marked as success as long as its protocol commands have been sent and parsed correctly. Otherwise, this request will be regarded as a failure, either re-transferred or skipped according to parsers' results. One request failure will not block the next requests to be transferred over the same connection as long as this connection is not broken. 

\subsection {Fetch/Prefetch Services}
SMURF-Cloud transfers the large-scale metadata using \textit{concurrency} and \textit{pipelining} and guarantees the reliability of the transfers in WAN. The details of sub-components and features are described below.


\subsubsection {Metadata Transferring via a Cluster of Fetch/Prefetch Services} \label{smurfcluster}

The dispatcher assigns the pending jobs to all available services in Round-Robin. One service will become available to process the next job as long as the service completes one fetch/prefetch job and sends back an ACK message to the queue dispatcher. When a service is terminated, the unacknowledged jobs will be re-dispatched to the rest of the available services. A fetch/prefetch service keeps at most one \textit{singleton connection} to the remote server and serves up to $C$ (pipelining capacity) fetch/prefetch jobs from the queue. If one established connection has been idle for a while, this connection will receive a TIMEOUT reply from a remote I/O node, and resources of this connection will be de-allocated from the service. Re-establishment of connection will be triggered by the next dispatch job and automatically handled by the Transferring Stream~\ref{transferstream}.



To fully exploit the computing power and network I/O bandwidth of the cloud cluster, the cloud backend deploys and launches multiple instances of fetch/prefetch services on a single cluster node. With $N$ services running in the cluster, the cloud establishes $N$ TCP connections to remote servers and transfers $N$ metadata requests concurrently. The cloud backend controls the \textit{concurrency} level by changing the number of fetch/prefetch service instances depending on the demand and the load. When the cloud deploys services across $M$ machines, the cloud service can transfer $N$ metadata requests from $M$ nodes that exceed the single machine's limitation and bottleneck. Meanwhile, the cloud service can tolerate the failure of ($M-1$) nodes. The instances of services running on the failed nodes will be redeployed to the other available nodes.

\subsubsection {Metadata Storage and Transfer Format}
On the cloud backend, the metadata is stored as a \{{\em key, value}\} pair in a NoSQL database, where {\em key} is the hash value of the resource path and {\em value} is the metadata content in JSON format (schemaless data structure). On the edge/fog node, the metadata resides in the memory cache as ProtocolBuffer~\cite{ProtocolBuffers} objects. When transferring the metadata between the system nodes, the metadata content is encoded into bytestreams by ProtocolBuffer.

The size of metadata can be as small as a few bytes or up to hundreds of megabytes. In our experiment, Figure~\ref{fig:s3filesdepthperdir} shows that a few directories can contain more than 400 thousand subfiles in the trace log. The overhead of encoding, decoding, and transferring such large metadata content between the continuum cache layers will severely degrade the system response time and increase the user-perceived latency. Thus, the cloud divides the large metadata object into fixed-size blocks and guarantees that each block object's size will not exceed the size limit. Very large-sized metadata will be stored as a bunch of metadata blocks, and these metadata blocks will form into a logic tree structure, where the blocks of partial subfiles will be stored at the bottom as the leaf nodes. A manifest of sub-blocks will be stored as part of the metadata to locate the metadata blocks, and any metadata block can be accessed by a uniform resource identifier (URI).

The metadata block is stored in a distributed storage; thus, the services can access and update the metadata blocks in parallel. Moreover, the pipeline and concurrency transferring of the metadata blocks significantly increase the end-to-end system throughput between the hierarchical layers. This also benefits the prefetching since once the metadata block has been received and decoded, its content can be immediately available to the cache instead of waiting to transfer and decode the original metadata. Thus, the prefetch predictor can adjust its prefetching decision on time based on the real-time prefetch results. SMURF system judiciously chooses the right block size to avoid storage space overhead and the cost of bandwidth on transferring. The effect of metadata block size selection on average fetch latency and memory usage is evaluated in Section~\ref{Evaluation}.

SMURF resolves the overwrite conflicts of metadata contents using the timestamp of files on remote I/O node as the ``version''. The underlying database clusters can guarantee the atomic read/write on the same metadata entry. Moreover, as long as the retrieval metadata's file timestamp is newer than what has been cached, it is safe to overwrite it. Otherwise, the retrieved metadata with the stale timestamp will be discarded. The service will return currently cached metadata content to the edge/fog service that requested this metadata retrieval.

\subsubsection {Directory Tree Structure Synchronization} \label {DELETE}
SMURF provides a way to maintain the metadata consistency between cloud and remote I/O nodes on the directory tree structure. It caches metadata content under the {\em key} of the request URL. If a folder has been renamed, deleted, or moved on the remote I/O node, then the subfolder metadata cached in SMURF will become dirty. If any metadata retrieval with force-refresh option has such an invalid path, the service will receive ``No such file or directory'' exception in the reply from remote I/O nodes. 

Once a fetch/prefetch service gets such an error from a remote I/O node, then SMURF cloud does {\em backtrace} synchronization to conservatively clean up the cached metadata under those invalid file paths. 
First, the fetch/prefetch service will try to read the currently cached metadata digest $D$. Then metadata retrieval on the invalid file path will cause the underlying transfer stream to return ``DELETE'' ErrorCode, which means this metadata has been cached under the invalid file path and should be marked as deletion. The atomic operation will compare and overwrite the ``DELETE'' status into current caching metadata $D'$ if $D$ is equal to $D'$. The comparison of metadata content digest guarantees that this service will mark ``DELETE'' status on the invalid caching metadata without overwriting the metadata content of another success update $D''$. Finally, if this deletion has been successfully populated into cloud DB, the deletion message will be sent to update all subscribed edge/fog servers which have fetched/prefetched on this invalid file path previously. Otherwise, the service will return the current caching metadata. Moreover, when deletion happens, the fetch/prefetch service will create a new fetch/prefetch request to do force-refresh on the parent file path and prefetch sublayer files (without force-refresh). 

Cloud will synchronize the parent file path's metadata content from remote I/O node and then prefetch 1-layer sub-folders under this newly updated metadata content. Without a force-refresh option on the subfiles prefetching is to maximumly reused local cache to avoid redundant force-refresh retrieval of the cached metadata. It is possible that the parent file path could also be invalid, then the fetch/prefetch service will repeat the above process to synchronize the metadata on the parent file path and increase the prefetch option on subfolder layers by 1, e.g., prefetch 2-layer (prefetchTTL=2). The cloud backend performs early-stop prefetch, which means such propagation of prefetching will terminate when a file path is valid or has not been cached yet.

\subsection {Distributed Continuum Caching and Prefetching Architecture}\label{continuum-cache}
The same metadata will be cached in the distributed layers-\{1,2,3\}, as shown in Figure~\ref{fig:usercase}. 
Our experimental analysis simulates the IoT network topology and assumes that the cloud has unlimited storage space, and the fog node can have a larger cache capacity than the edge server. The system consists of the distributed continuum cache from cloud to an edge server, where the cloud caches and stores all fetch/prefetch metadata, the fog node caches partial cloud metadata, and the edge server only caches a small subset of the fog metadata. The prefetch predictors can be installed on the edge server and the fog node with the judicious parameters to retrieve the locality metadata into each layer's local cache.
Usually, our system will conduct more aggressive prefetching between the fog node and the cloud to mask the high latency in WAN. 
The prefetching between the edge server and the fog node is more conservative by considering that the edge server connects to the fog node in LAN (usually wireless connection) and the storage is always limited on the edge server. 

When the optional fog node (layer-2) has been deployed between the edge server and cloud, this edge server will fetch/prefetch metadata from the fog node's local cache, which can be denoted as $F_{edge}$ and \{$P_{edge}$\}. The fog node can send the cached metadata back to the edge server or forward the cache miss fetch request $F_{edge}$ and prefetch requests \{$P_{edge}$\} to the cloud. The cache miss fetch request $F_{edge}$ can cause the fog node's prefetch framework (more details in~\ref{prefetchframework}) to consult its prefetch predictor on the aggressive prefetching \{$P_{fog}$\}. Thus there would be overlapping prefetch requests between \{$P_{edge}$\} and \{$P_{fog}$\} requests; however, the wait-notify queue (discussed in~\ref{wait-and-notify}) will de-duplicate the overlapping prefetch requests to send them to the cloud, and the fog node will send back the edge server's requested prefetch metadata \{$P_{edge}$\} upon completion.

\subsubsection {Layer Server's Request and Response Multiplexing}
\label{wait-and-notify}
We design and implement a \textit{wait-and-notify queue} to efficiently send and receive messages between the edge server and the cloud. The wait-and-notify queue consists of a sender thread and a receiver thread. Multiple worker threads can enqueue the requests and wait for the notification of completion concurrently. The sender thread allocates a unique context locally for each enqueued request and sends requests to the remote layers. The receiver thread extracts the context from the response and then uses this context to notify and wake up the waiting worker threads. Especially during one request $R$'s sending and receiving, the similar queuing requests will be de-duplicated without sending them to the cloud, and their worker threads will wait for the completion of $R$. The de-duplication is executed on the edge/fog layer to ease the cloud's computing overhead and save the network bandwidth. The queue can also be configured to be in a ``nowait'' mode, which means worker threads do not wait for the completion. The wait-and-notifying queue mechanism exhibits high performance in multiple threading environments. Message sending and receiving are designed to be interleaved between multiple threads. Moreover, the queue has been implemented based on a non-blocking queue, where {\em compare and swap (CAS)} strategy has been applied to improve the concurrent performance without the blocking synchronization. Note that the order of requests sending is not necessarily synchronized with the receiving order of the responses. 


\subsection{Prefetch Framework}\label{prefetchframework}

SMURF employs a generic prefetch framework to apply the configurable prefetch predictor on edge/fog service. Users and system admins can easily configure and customize prefetch schemes for different types of applications. In this prefetch framework, each fetch request will be sent to the prefetch predictor to analyze and build a prefetch correlation relationship. For each fetch request, the prefetch framework maintains a cache miss counter and its metadata content in the cache with Least Recently Used (LRU) replacement policy. The cache miss counter denotes how many cache misses are on this request. The cache object can be evicted using the LRU replacement algorithm when the cache is full, and new metadata needs to be put into the cache. When the cache miss counter's value exceeds the threshold, the prefetch framework will consult the prefetch predictor for the potential prefetching candidates and execute aggressive prefetching on this request's correlation candidates. 
The prefetch framework checks whether each prefetch candidate exists in the current local cache. If there is a cache miss on this candidate, the prefetch framework will pack and send a prefetch request with the information (e.g., URI and priority). Prefetch framework does not maintain the cache miss counter for all the history requests since the essence of the LRU cache replacement algorithm is based on temporal locality, and the cache miss information of the coldest request will be replaced and cached out to reflect the temporal access locality and also save the memory usage.

The edge/fog node can set the value of the threshold in the prefetch predictor property file. This threshold is an essential factor to affect the frequency of prefetch on correlation files. If the threshold value is set to a smaller value, then the prefetch predictor will be too sensitive to detect the cache miss on a file path and aggressively prefetch correlation files. However, the too big threshold will cause the prefetch predictor to react to the cache miss without sending prefetch requests slowly. The threshold is tuned by the analysis of the trace log. 

\subsection {Semantic Locality Prefetch Predictor}

SMURF uses a novel prefetch predictor based on the directory semantic locality. The predictor uses a history window to predict the semantic locality of the trace log. This history window with the fixed window size stores the unique file path into segments. For one input file path, the predictor will find out the pattern of ``A ? B'' with the maximum matching number inside the history window, where ``A'' stands for the common prefix, ``?'' stands for one mismatch segment and ``B'' is the suffix, and sometimes this suffix can be empty. If the matching number exceeds the threshold, the predictor sends back this detected pattern to predict that the fetching file paths will follow this access pattern in the near future.

When a request on a file path $f$ causes a local cache miss, the predictor will check whether its pattern file path $f_{p}$ object is cached or not. If the pattern file path object has not been cached, then the predictor will create an object of this pattern file path, put it into the local cache, and set the counter's value to one. If the pattern file path object has been cached, then the predictor will increase the cache miss counter by one. When this cache miss counter exceeds threshold $T$, the predictor decides to prefetch the correlation files of the pattern file path $f_{p}$ and set the miss counter to zero.

If the pattern file path object has already been cached, the predictor will iterate the list of metadata of each subfile path $f_{si}$ in the cache and send the prefetch requests of all cache missed subfile paths. In the prefetch request, the predictor can associate with the value of prefetch TTL, configured (by default, the value is 0) in the prediction property file. The number of prefetch TTL is to denote how many layers of subfiles to prefetch. Theoretically, the predictor can set an arbitrary large number to the value of prefetch TTL. However, upon completion of the prefetch on a file path, the queue system will automatically decrease the value of prefetch TTL by one and recreate a new prefetch request for each subfile and then re-queue those requests with the lower priority until TTL degrades to 0. In the competition of large-scale prefetching requests, higher priority prefetch requests will be given precedence and always preempt available prefetching services. It could be a large amount of lowest priority prefetch requests in the queue system, which can never be served or completed in a period and will be finally reclaimed and destroyed by queue cleaning.

Semantic locality predictor can effectively match the workload pattern on the semantic directory tree structure, primarily when most of the subfiles under a common hotspot parent file path are randomly accessed once or very few times. When the data access sequence does not exhibit strong locality behavior under a common file path, semantic locality predictor can set a higher threshold to prevent redundant miss-prefetching effectively. Semantic locality predictor can potentially prefetch a large amount of metadata in WAN. However, our system has been optimized to support the massive scale of metadata transfer with lower average latency in WAN, which can fit the semantic locality predictor requirements.

\begin{figure}[t]
\centering
\includegraphics[width=0.98\columnwidth]{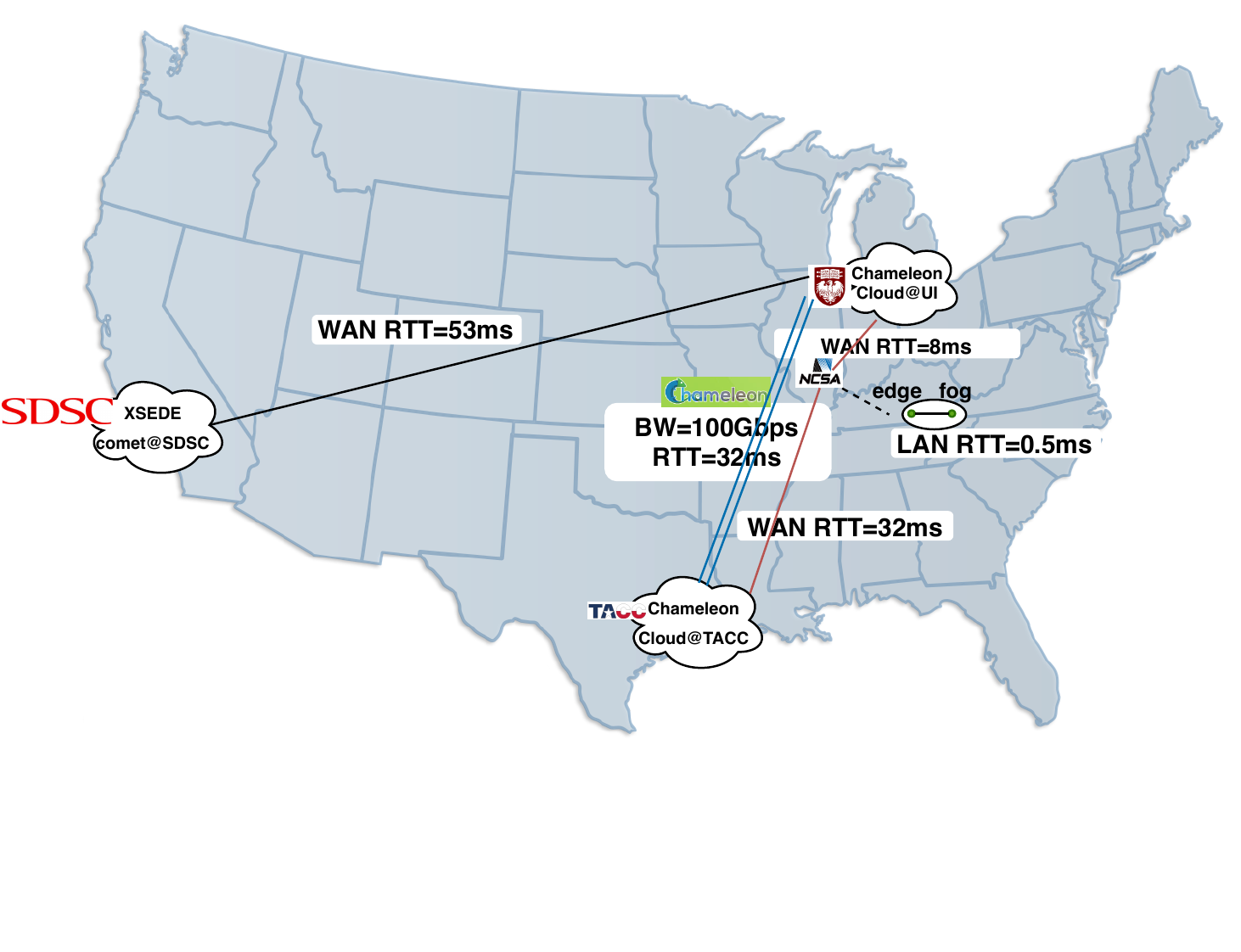}
\vspace{-16mm}
\caption{Network map of the experimental testbed.}
\label {fig:geo-network-map}
\end{figure}

\begin{table}[t]
\caption{Specifications of the Cloud/Edge/Fog nodes in the experiments.} \label{tab:machine-specs}
\begin{centering}
\footnotesize
\begin{tabular}{ |c|c|c|c|}
\hline
{\bf Specs} & {\bf Edge/Fog} & {\bf Cloud} & {\bf Remote IO}\\
\hline
\hline
CPU & \shortstack{Intel Core\\i7-2600} & \shortstack{Intel Xeon\\Gold 6126} & \shortstack{Intel Xeon\\E5-2650 v3}\\
\hline
RAM & 32 GB & 187 GB & 62 GB \\
\hline
Disk & 80 GB & 210 GB & 350 GB \\
\hline
OS & Ubuntu 16.04 & Ubuntu 16.04 & Ubuntu 16.04\\
\hline

\# of nodes & 5 KVM & 5 Bare metal& 1 Bare metal\\
\hline

\end{tabular}
\label{tab:system-machine-spec}
\end{centering}
\end{table}

\section {Evaluation} \label {Evaluation}


We conduct our experiments over Yahoo! Hadoop grid trace logs from Yahoo! Webscope dataset~\cite{yahoo-webscope}. This trace consists of more than 20 Million continuous daily metadata operations of the Hadoop name node throughout the year 2010. Geographical locations of the servers used in our experiments and the network specifications between them are presented in Figure~\ref{fig:geo-network-map}. The system settings and configuration are shown in Table~\ref{tab:system-machine-spec}. To simulate the heterogeneous remote I/O nodes, we setup FTP and iRODS servers at the Chameleon-TACC site and installed Globus Toolkit 6.0~\cite{globus-toolkit} to configure the SimpleCA and GSIFTP server for the GSIFTP metadata transfers. The Minio~\cite{minio} server is installed to simulate the Amazon S3 object storage service. In all experiments, SMURF protocol libraries are installed only on SMURF-Cloud, deployed on the Chameleon-UC bare-metal cluster. SMURF's fetch/prefetch services are launched as Docker~\cite{boettiger2015introduction} services and managed under the Docker orchestration tool. The edge and fog clusters are deployed into a Kernel-based Virtual Machine (KVM) cluster and configured with limited computing resources for more realistic experimental evaluation.

\begin{figure}[t]
\includegraphics[keepaspectratio=true,angle=0,width=1\columnwidth]{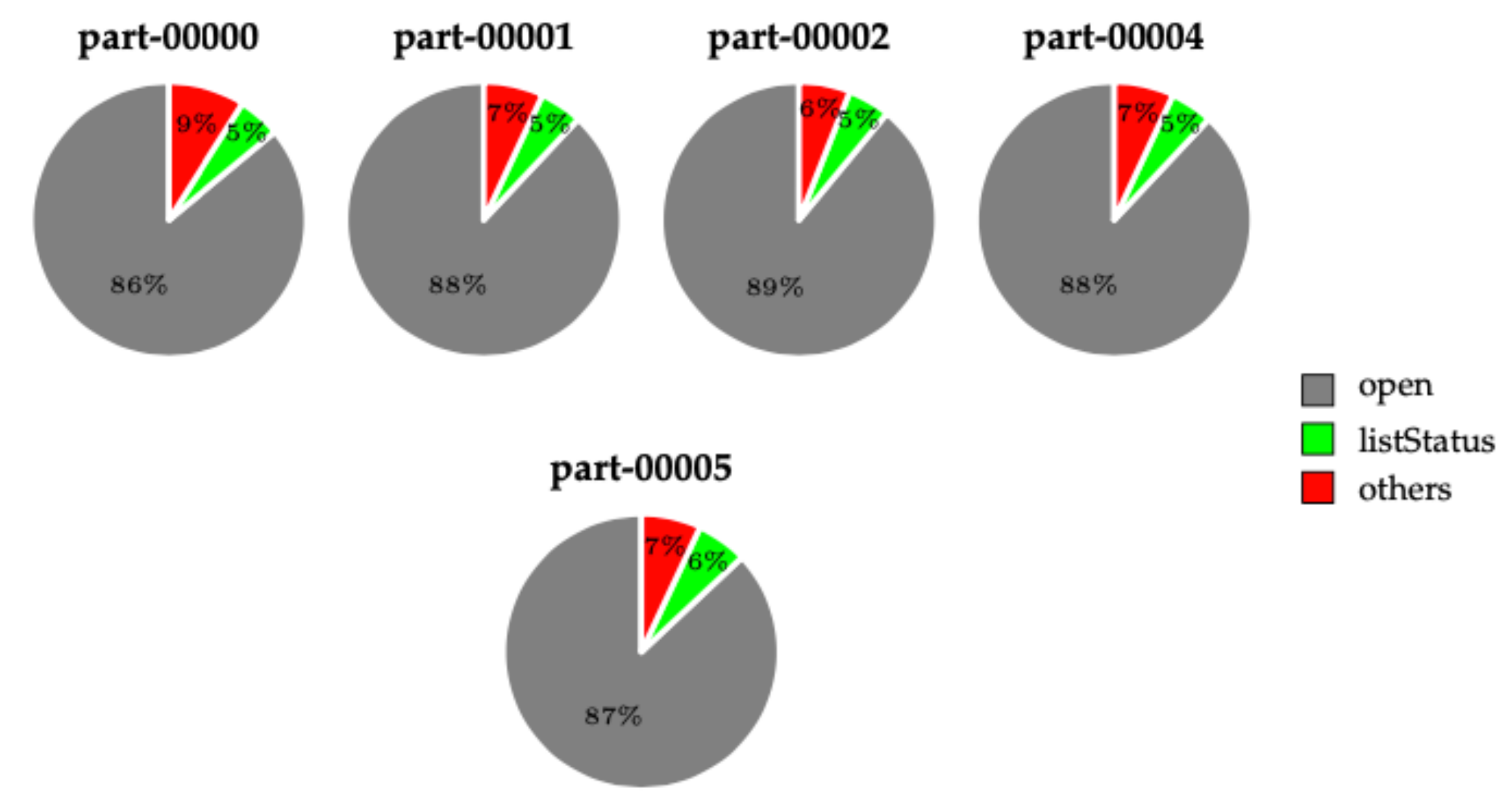}
\caption{The distribution of the metadata operations in Yahoo! Webscope Hadoop traces.}
\label{fig:metadata-operation-types}
\vspace{0.5mm}
\end{figure}

\begin{table}[t]
\caption{Yahoo! Hadoop log's `list' command statistics.} \label{tab:s3-list-stats}
\begin{centering}
\footnotesize
\begin{tabular}{ |c|c|c|c|}
\hline
{\bf Log Name} & {\bf \# of list cmds} & {\bf unique file path}  & {\bf histogram=1} \\
\hline
\hline
part-00000 & 4,750,645 & 49.72\% & 92.6\% \\
\hline
part-00001 & 4,090,678 & 62.31\% & 92.98\% \\
\hline
part-00002 & 3,732,058 & 62.52\% & 92.33\% \\
\hline
part-00004 & 3,895,900 & 62.77\% & 91.85\% \\
\hline
part-00005 & 4,148,414 & 54.23\% & 92.76\% \\
\hline

\end{tabular}
\end{centering}
\end{table}

\subsection {Trace File System Directory Tree Reconstruction}
In trace logs, file path $f$ always associates with types of operations, e.g., $open$, $ls$, $delete$, etc. Metadata read operations are directory tree idempotent (e.g., $open$ and $ls$) and will not change the directory tree structure on trace file system. The write operations (e.g., $mkdir$, $rename$ and $delete$) can change trace file system directory tree dynamically. 
Yahoo! Webscope dataset encrypts each segment of the file path into $27$ bytes string. Thus the approximate directory tree size of each Hadoop trace log on disk will be more than 250GB. 

\begin{figure}[t]
\begin{tabular}{cccc}
 \includegraphics[keepaspectratio=true,angle=0,width=0.48\columnwidth]{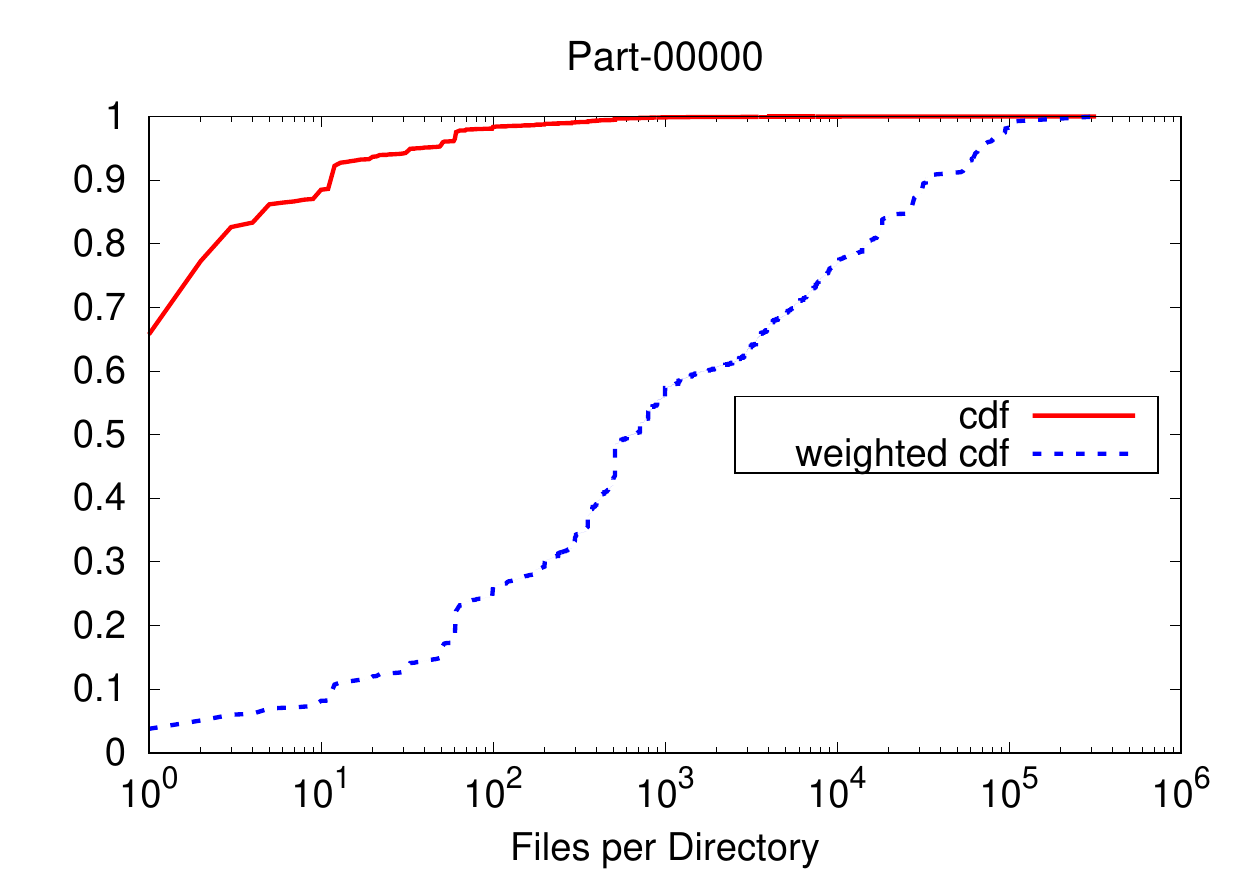}
 \includegraphics[keepaspectratio=true,angle=0,width=0.48\columnwidth]{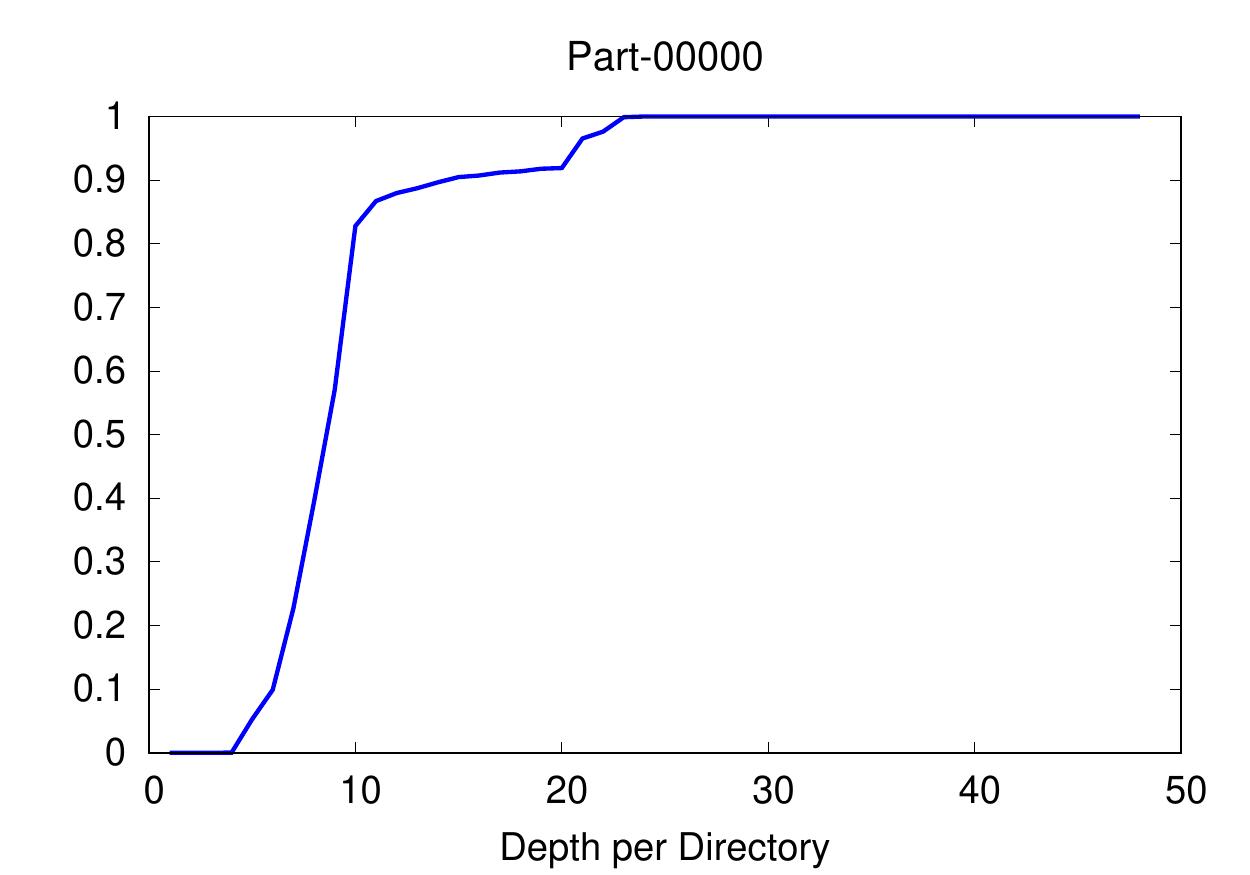} \\
  \includegraphics[keepaspectratio=true,angle=0,width=0.48\columnwidth]{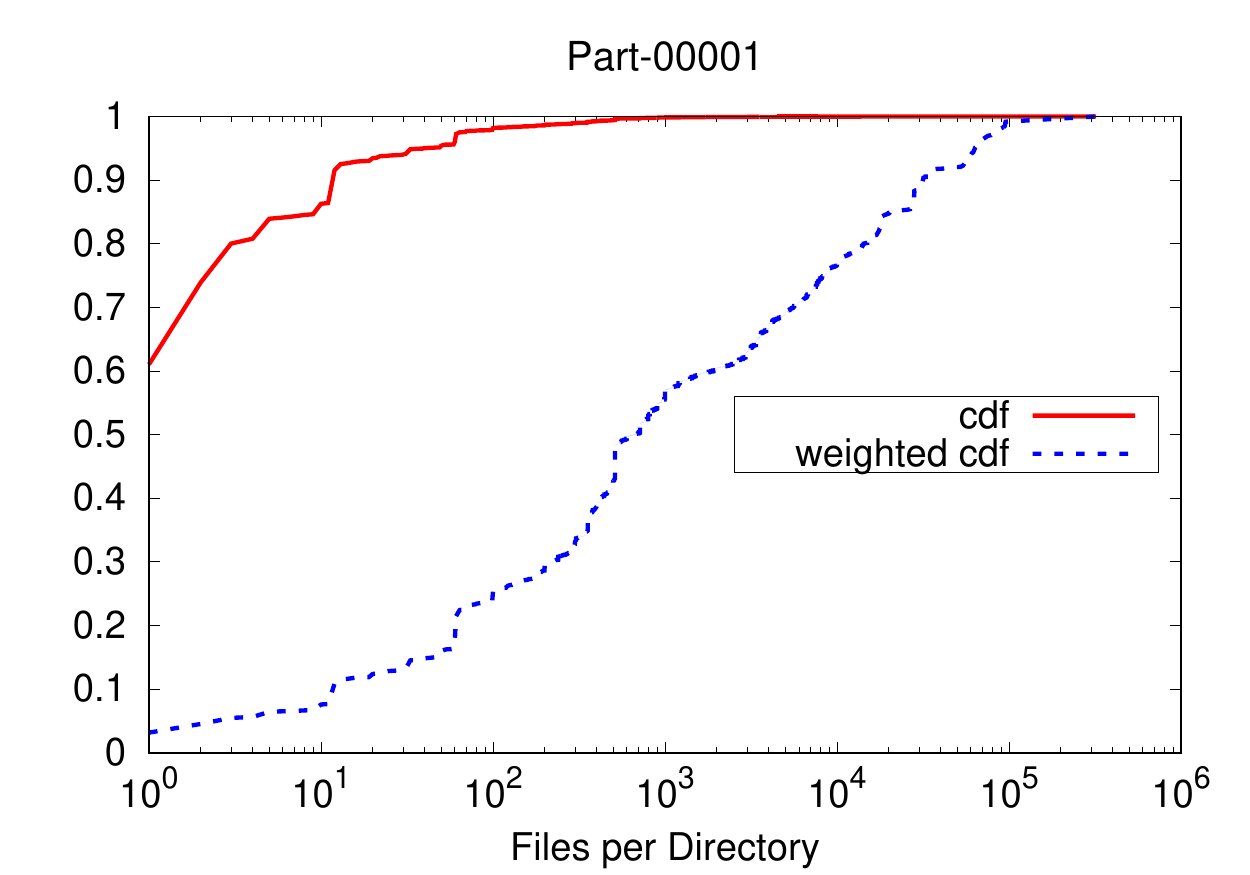}
 \includegraphics[keepaspectratio=true,angle=0,width=0.48\columnwidth]{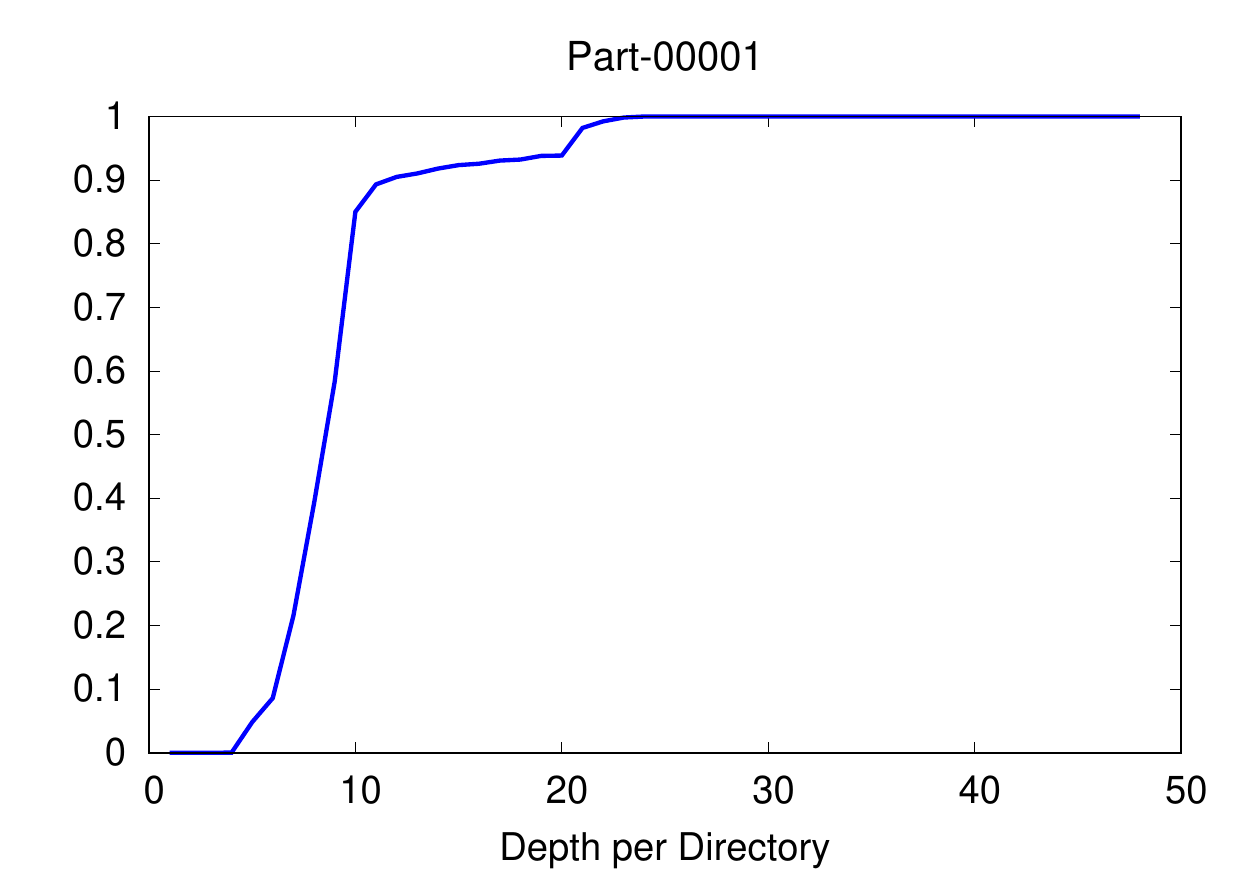} \\
   \includegraphics[keepaspectratio=true,angle=0,width=0.48\columnwidth]{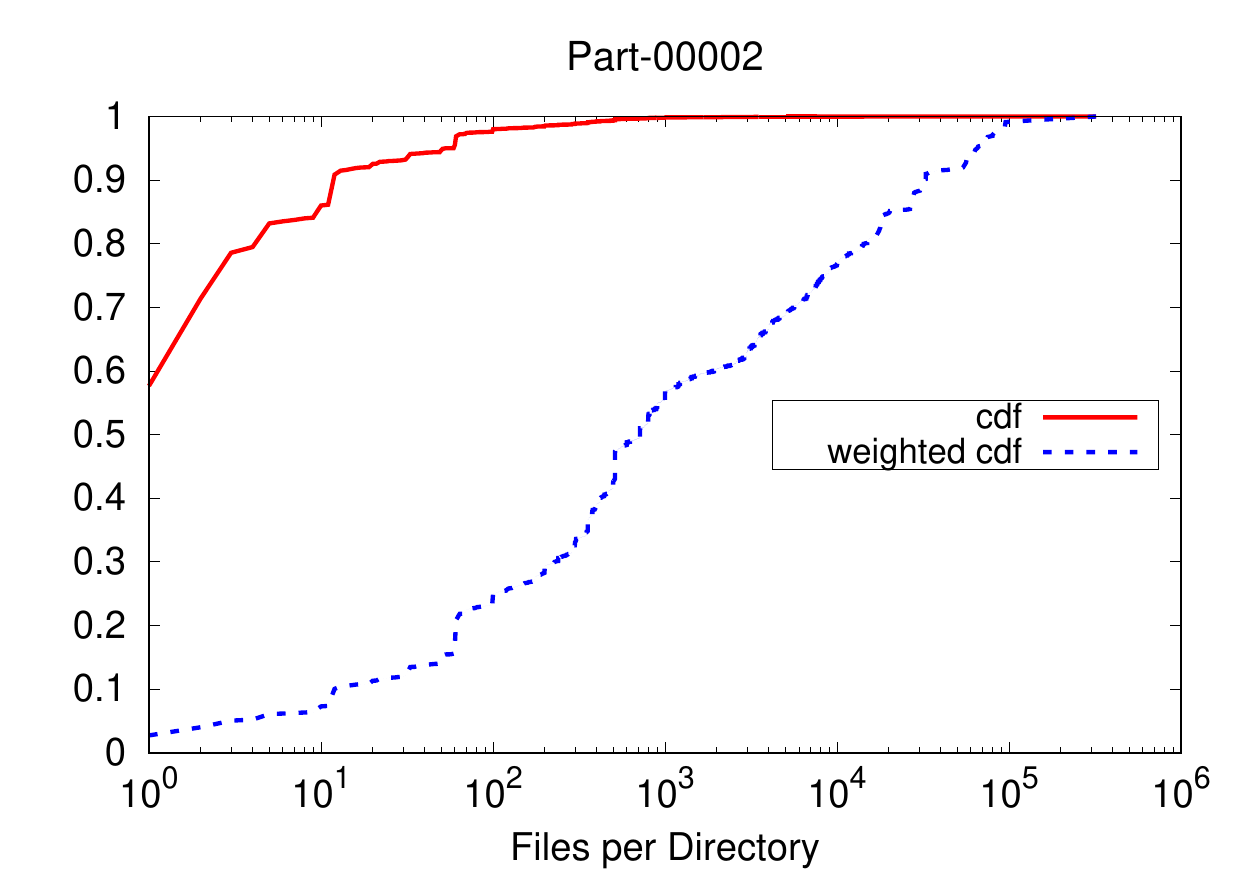}
 \includegraphics[keepaspectratio=true,angle=0,width=0.48\columnwidth]{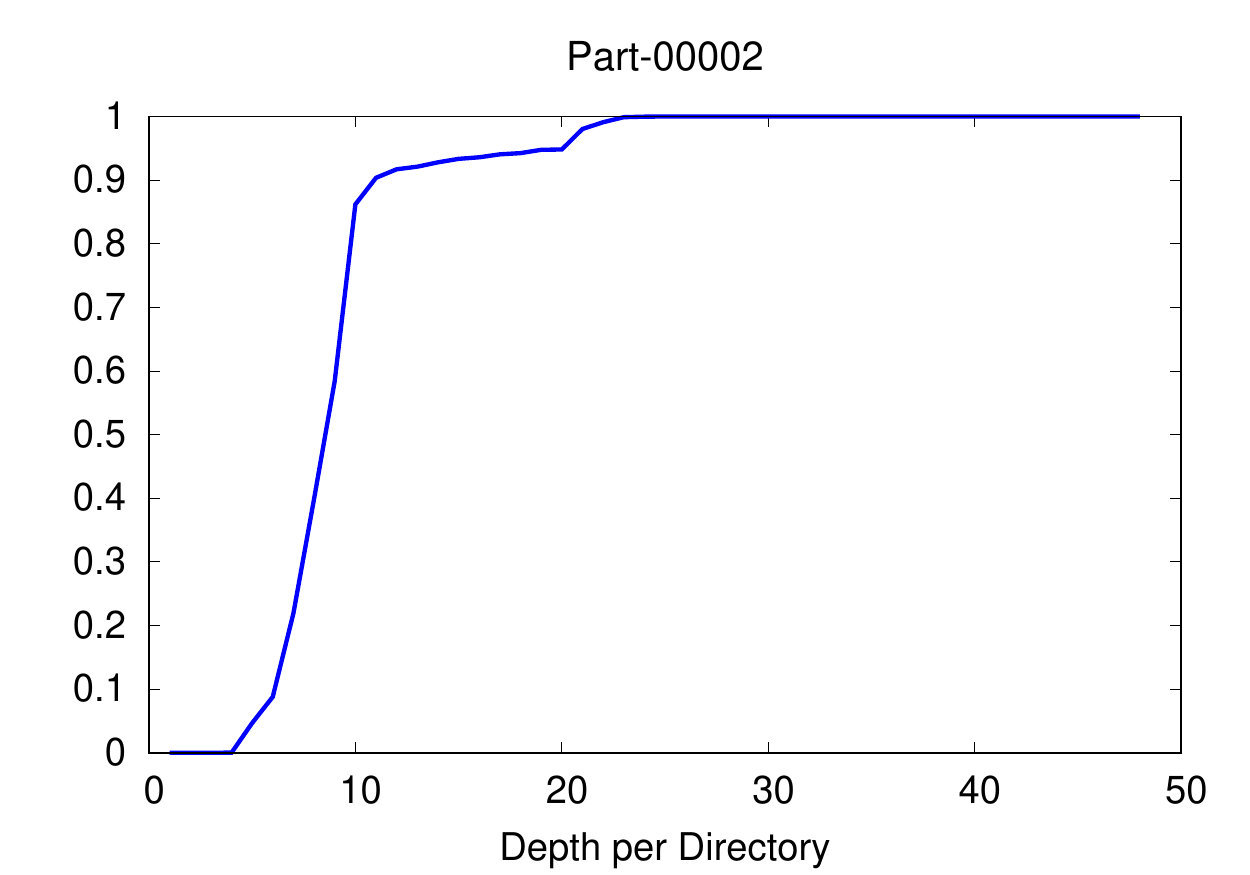} \\
   \includegraphics[keepaspectratio=true,angle=0,width=0.48\columnwidth]{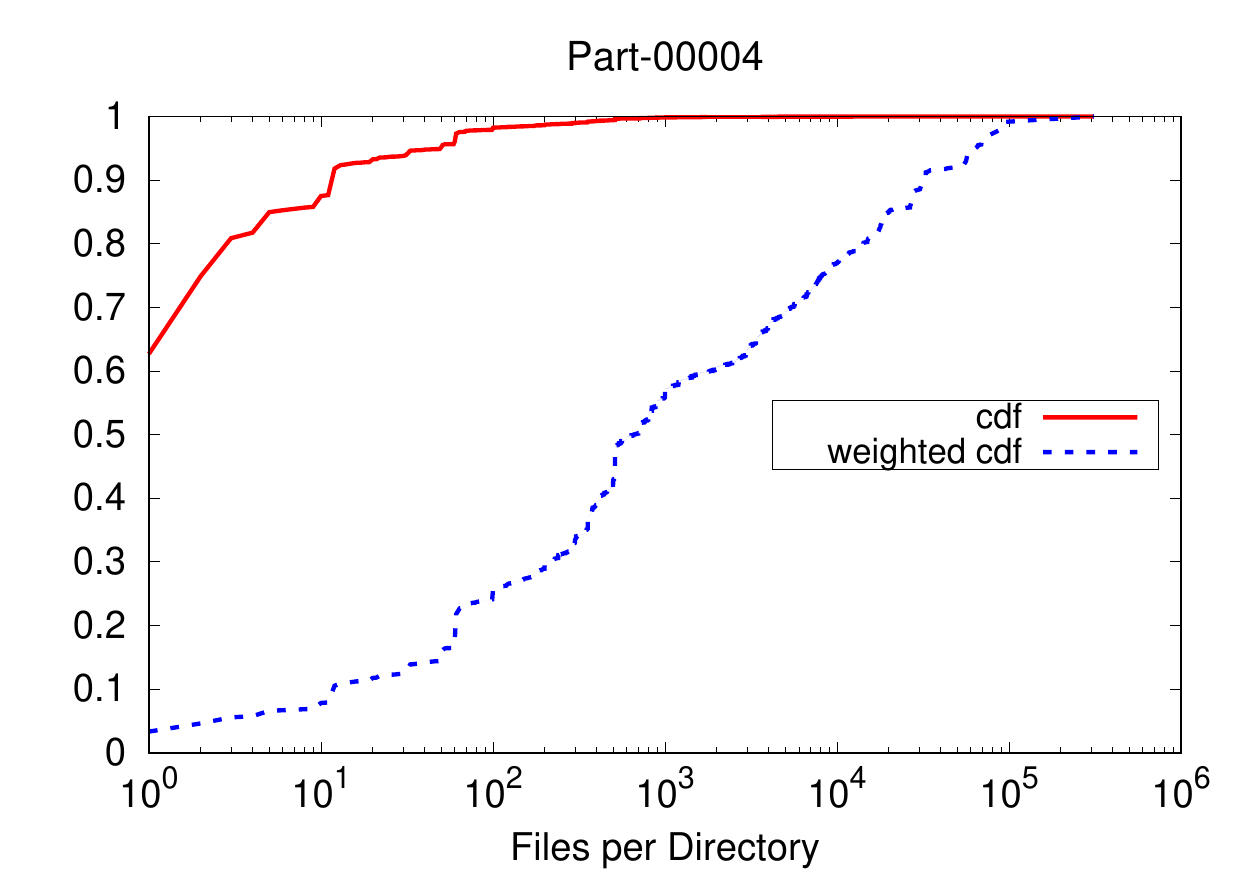}
 \includegraphics[keepaspectratio=true,angle=0,width=0.48\columnwidth]{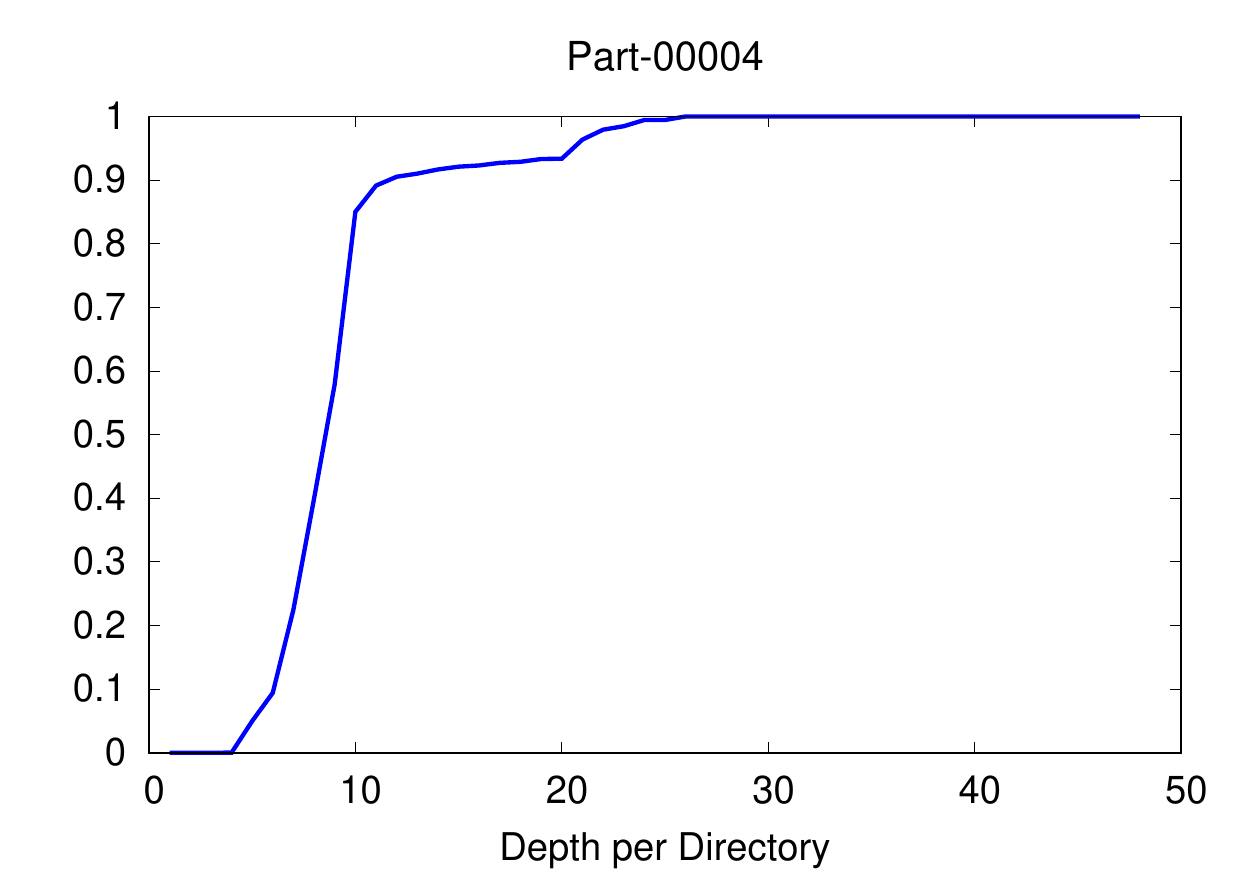} \\
    \subcaptionbox{}
    {\includegraphics[keepaspectratio=true,angle=0,width=0.48\columnwidth]{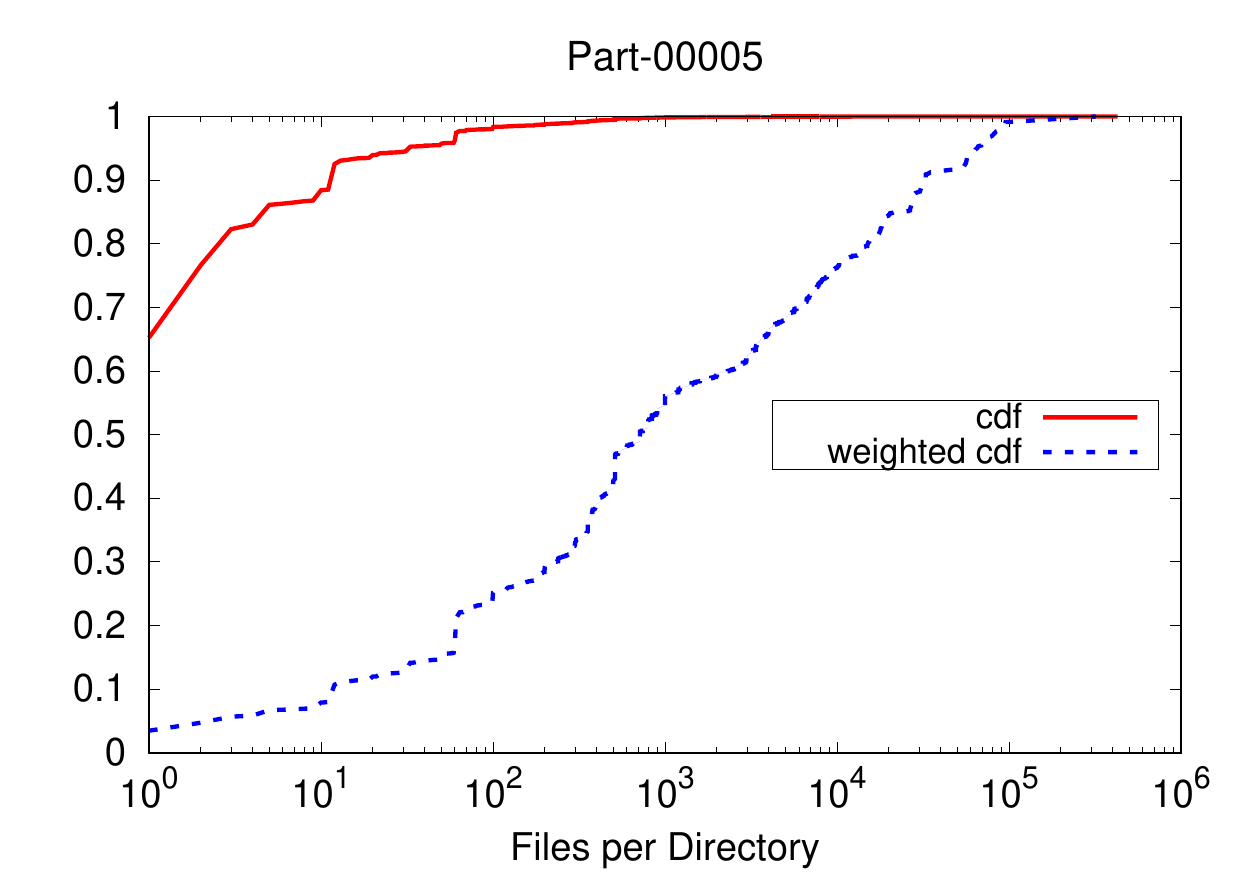}}

    \subcaptionbox{}
    {\includegraphics[keepaspectratio=true,angle=0,width=0.48\columnwidth]{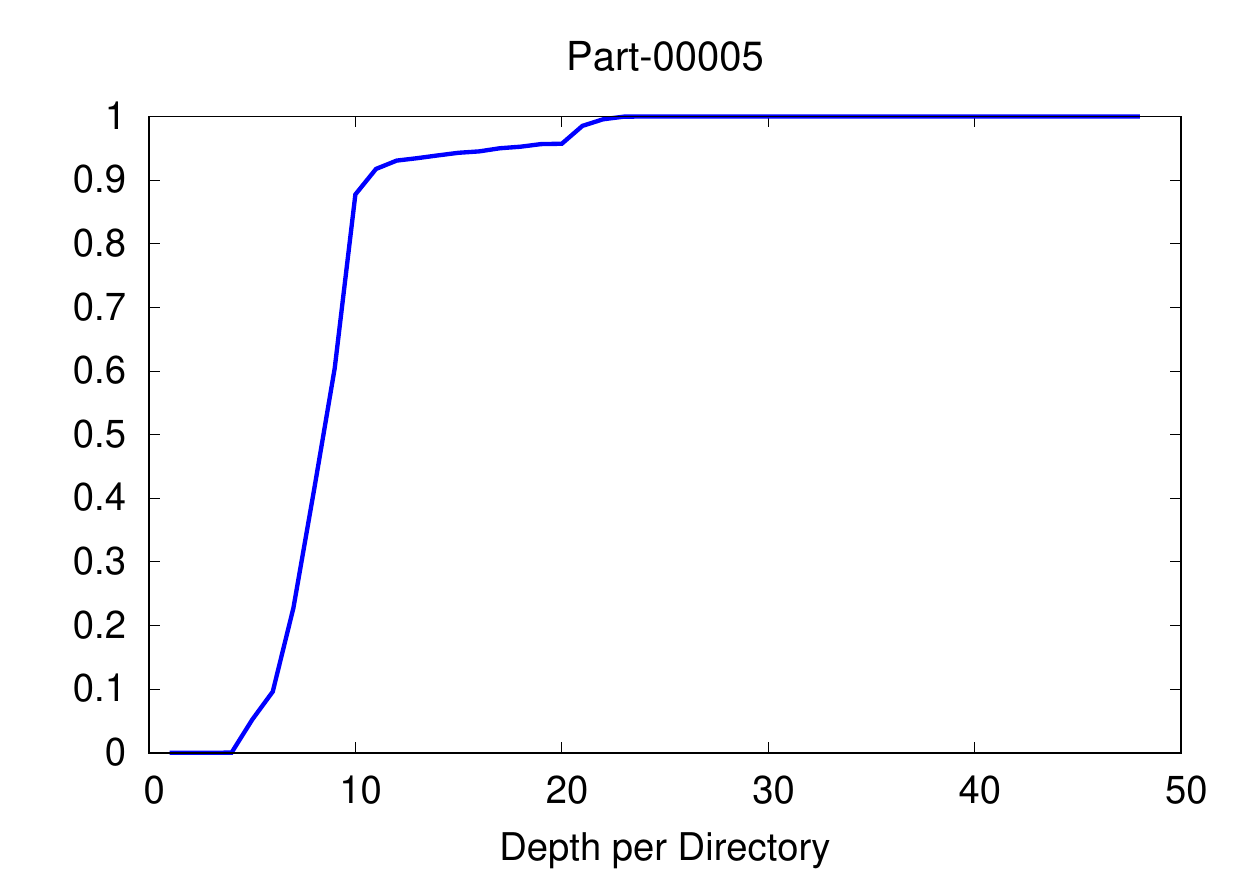}} \\

\end{tabular}
\caption{Trace file system statistics in Yahoo! HDFS: (a) CDF and Weighted CDF of files per directory; (b) Distribution of files by directory depth.}
\label{fig:s3filesdepthperdir}
\vspace{-4mm}
\end{figure}

We extract file paths from all types of operations for each audit log and construct them on the disk. This is the approximate emulation of trace file system directory tree structure in our prediction experiments. Figure~\ref{fig:s3filesdepthperdir} shows the cumulative distribution of directories by the number of files they contain and files by directory depth. This reconstructed file system's shape is flat: millions of files (nearly 90\%) reside in the directories with a depth between 5 and 10. Most of the directories (around 95\%) contain only a few files, and the majority of files (around 75\%) are stored under a small portion of directories (about 3\%), each of which contains more than hundreds of files and even up to hundreds of thousands of files.

We extract requests containing the ``listStatus'' command from Yahoo! Webscope Hadoop audit trace logs in our experiments. In table~\ref{tab:s3-list-stats}, we statistically analyze the histogram of distinct file path in ``listStatus'' command. The histogram results show very skew access of ``listStatus'' metadata operation in Hadoop audit log: among the total number of four million ``listStatus'' operations, there are 50\%-62\% of unique file paths. The majority (92\%) of unique file paths have been accessed only once, and only 8\% of unique file paths contribute nearly half of total ``listStatus'' metadata operations. This skew access to behavior can cause prefetch predictors based on historical access sequence abysmal prediction rate. Their prediction rate is almost the same as that of LRU cache since most of their prefetch candidates are from history requests, but they will never appear again in the next ``listStatus'' requests, and the most frequent file paths will reside in the memory by the cache replacement policy.

\begin{figure}[t]
\begin{tabular}{cccc}
 \includegraphics[keepaspectratio=true,angle=0,width=0.48\columnwidth]{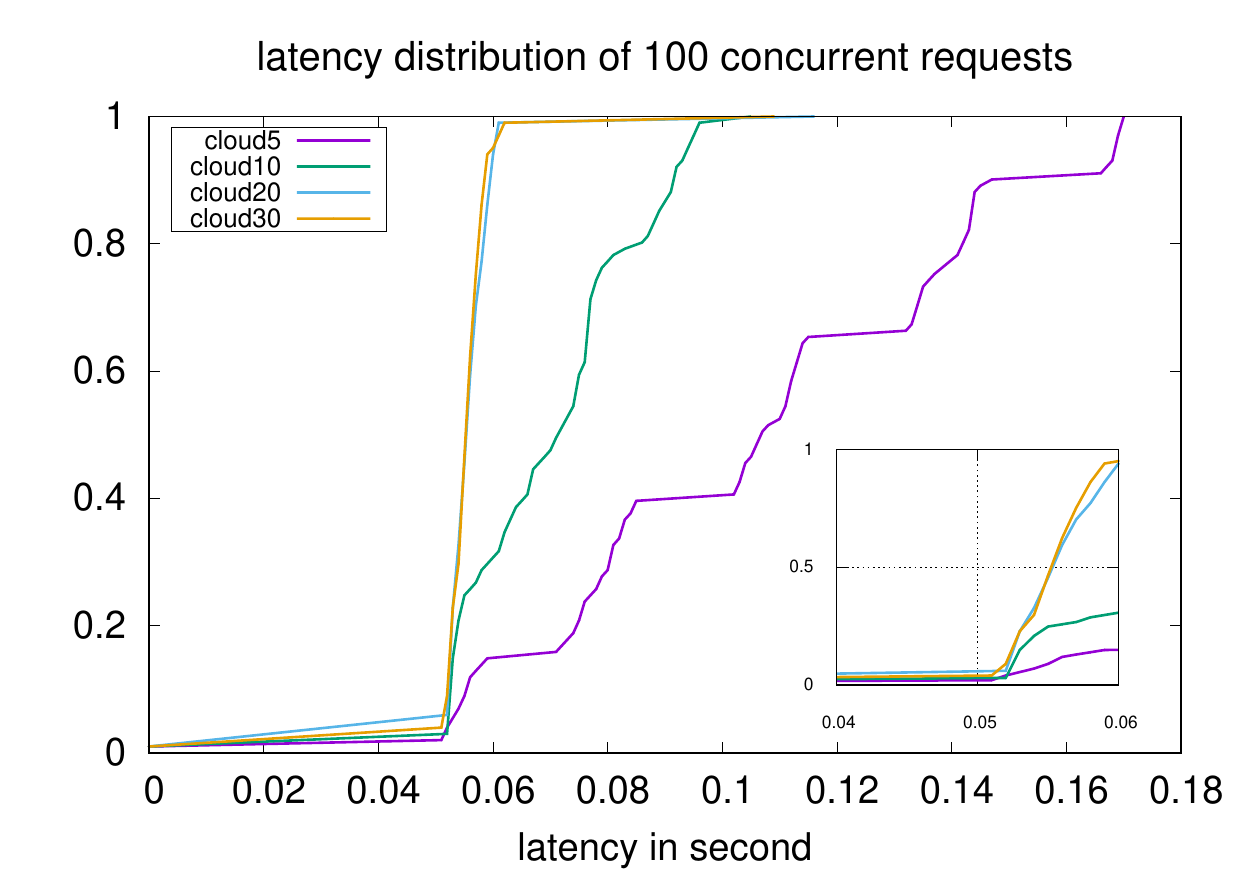}
 \includegraphics[keepaspectratio=true,angle=0,width=0.48\columnwidth]{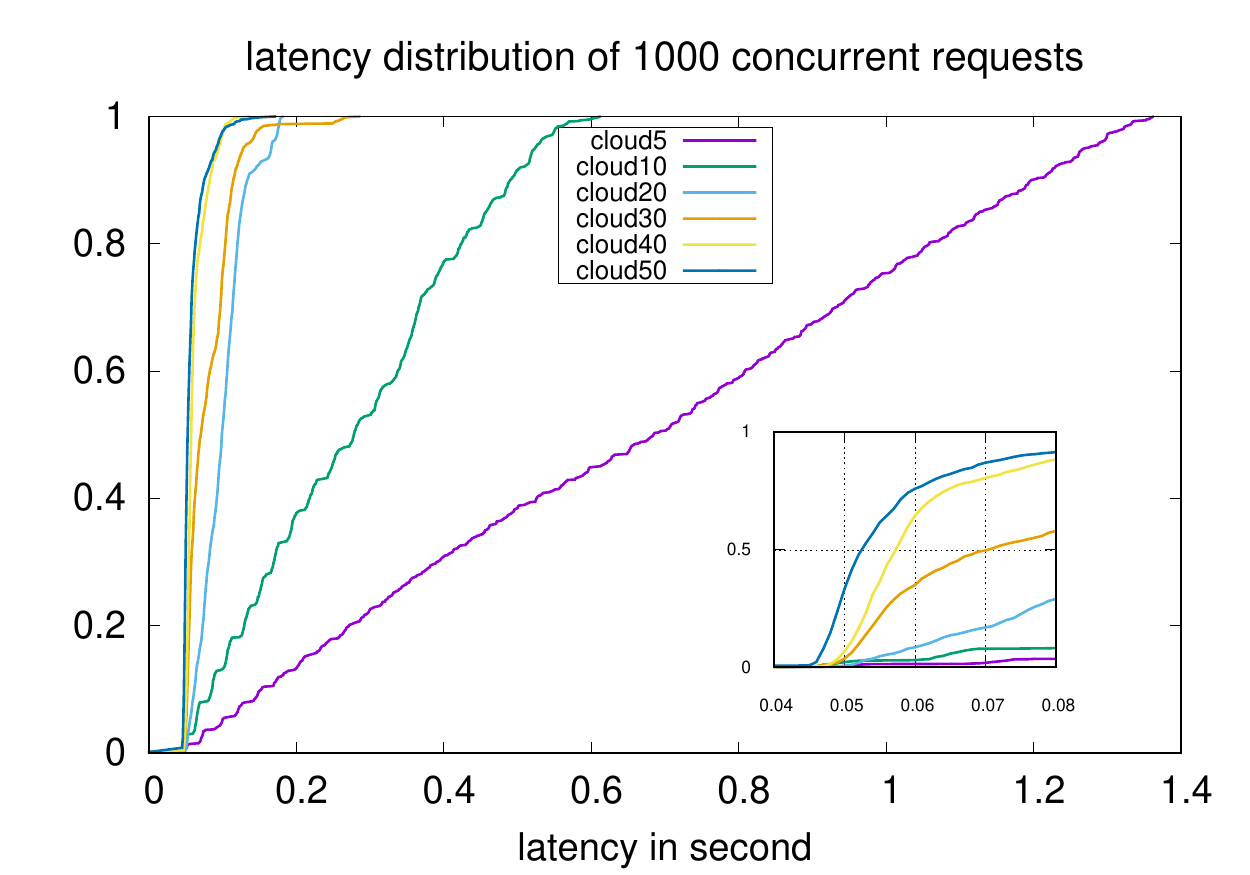} \\
\end{tabular}
\caption{Concurrent fetch latency distribution.}
\label{fig:multivmsdistribution}
\vspace{-4mm}
\end{figure}

\subsection {Scalability of Fetch Services}

To emulate concurrent metadata transferring performance from remote servers, we use Yahoo! Cloud Serving Benchmark (YCSB)~\cite{cooper2010benchmarking} to continuously send a large number of distinct requests from the client to the SMURF system and evaluate the latency distribution with the different number of cloud fetch services. All the requests will be transferred along the edge-cloud I/O path with around 40 ms accumulated RTT. In this experiment, we turn off the caching and prefetching effects in the testbed and configure the edge cluster with 16 fetch services, and set the value of each cloud fetch service's pipeline capacity to be 5. Figure~\ref{fig:multivmsdistribution} shows the latency measured on the edge cluster and demonstrates that the latency distribution curves of 100 and 1000 concurrent requests with five cloud fetch services are almost linear due to the queueing effect of cloud, which means five cloud fetch services are not sufficient enough to scale the number of concurrent requests.
Thus, with more number of services in the cloud, 
most of the requests can be made concurrently, and the latency of the majority of requests is within the small range between 40 ms and 80 ms. 

\subsection {Scalability of Prefetch Services} \label{heterogeneous-prefetch}

\begin{figure}[t]
\begin{tabular}{cccc}
    \includegraphics[keepaspectratio=true,angle=0,width=0.48\columnwidth]{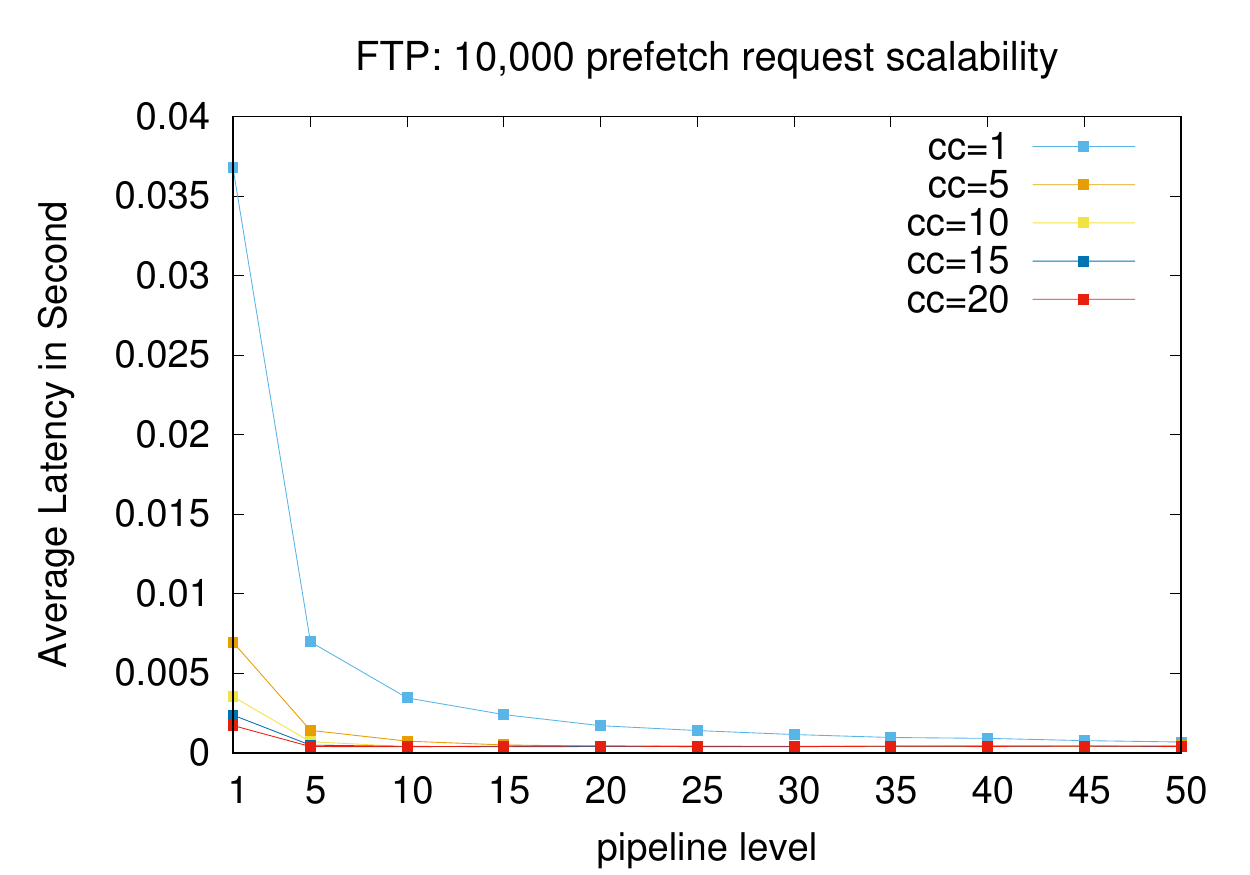}
 \includegraphics[keepaspectratio=true,angle=0,width=0.48\columnwidth]{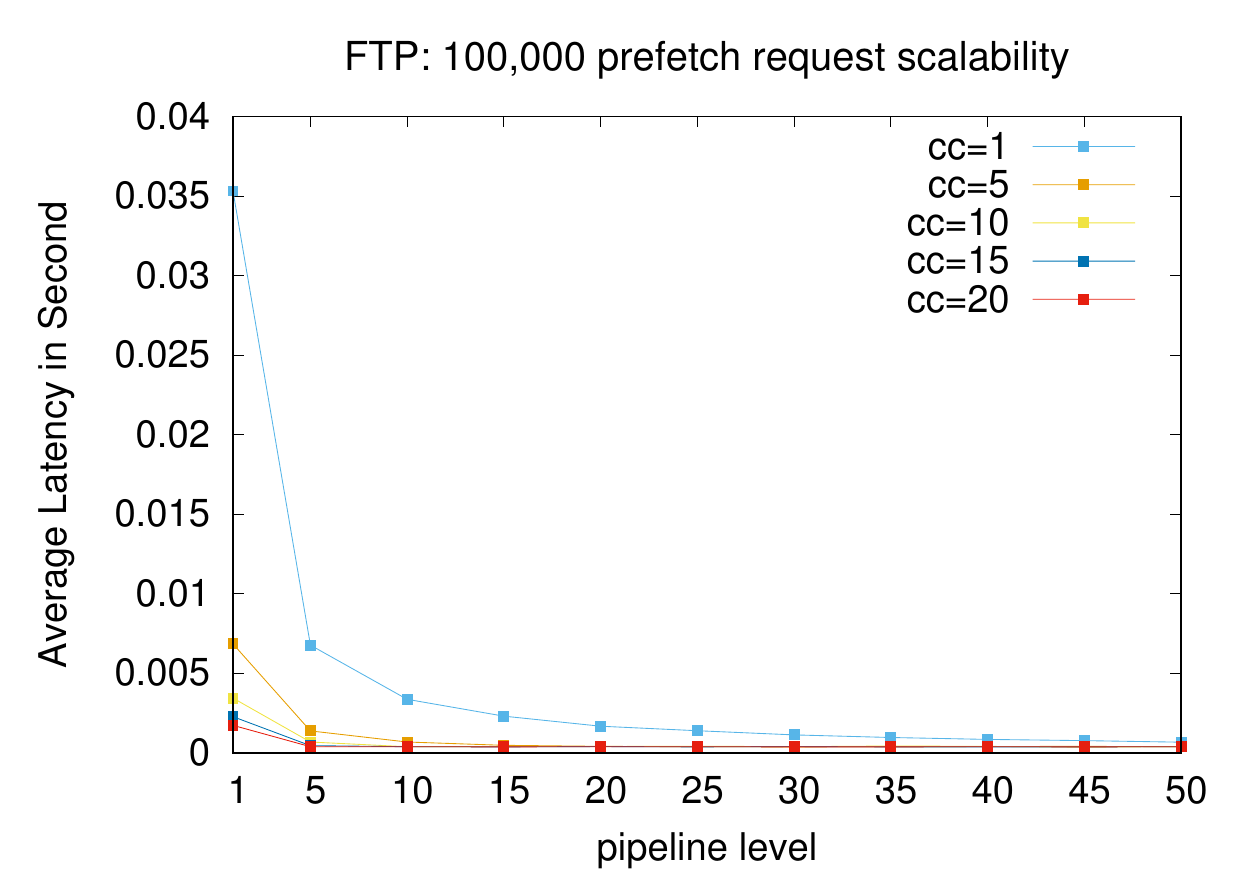} \\
    \includegraphics[keepaspectratio=true,angle=0,width=0.48\columnwidth]{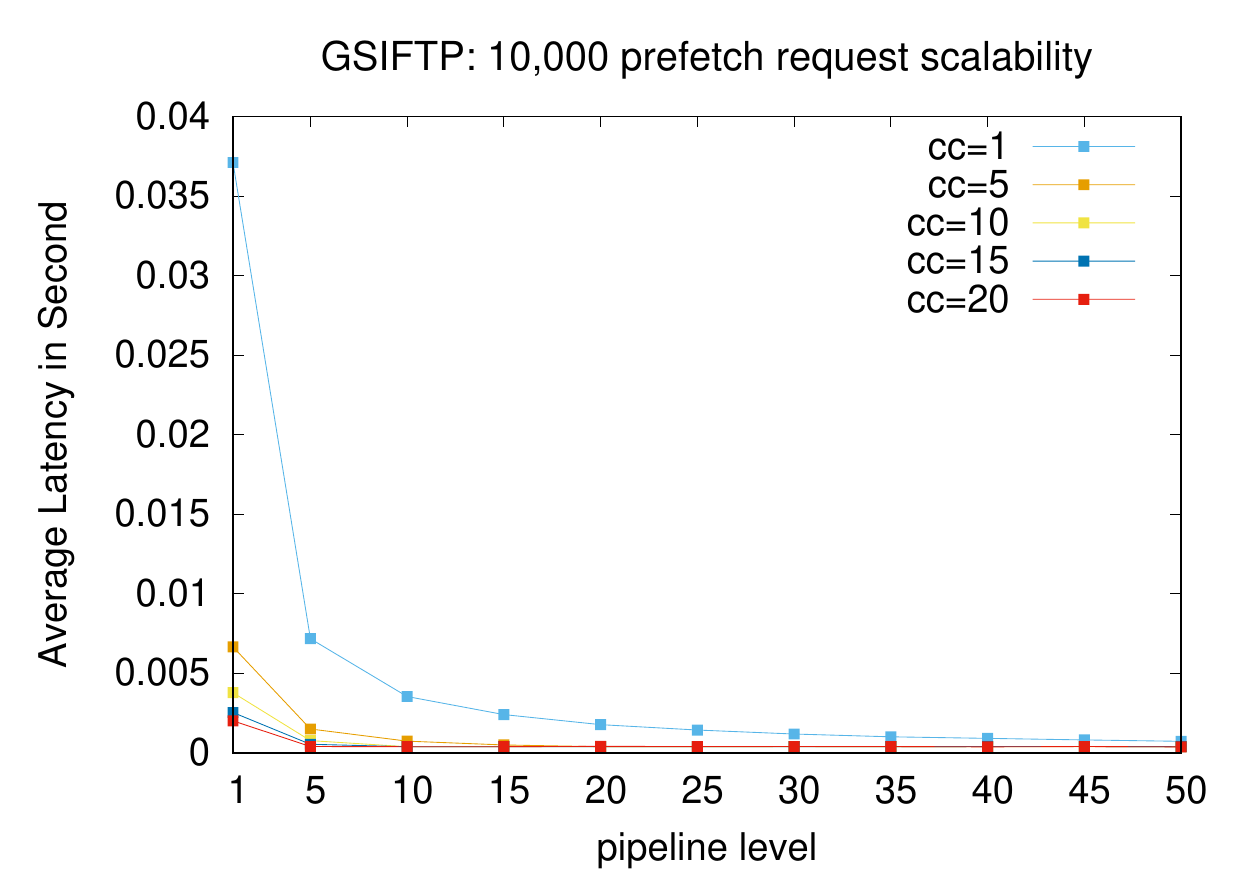}
 \includegraphics[keepaspectratio=true,angle=0,width=0.48\columnwidth]{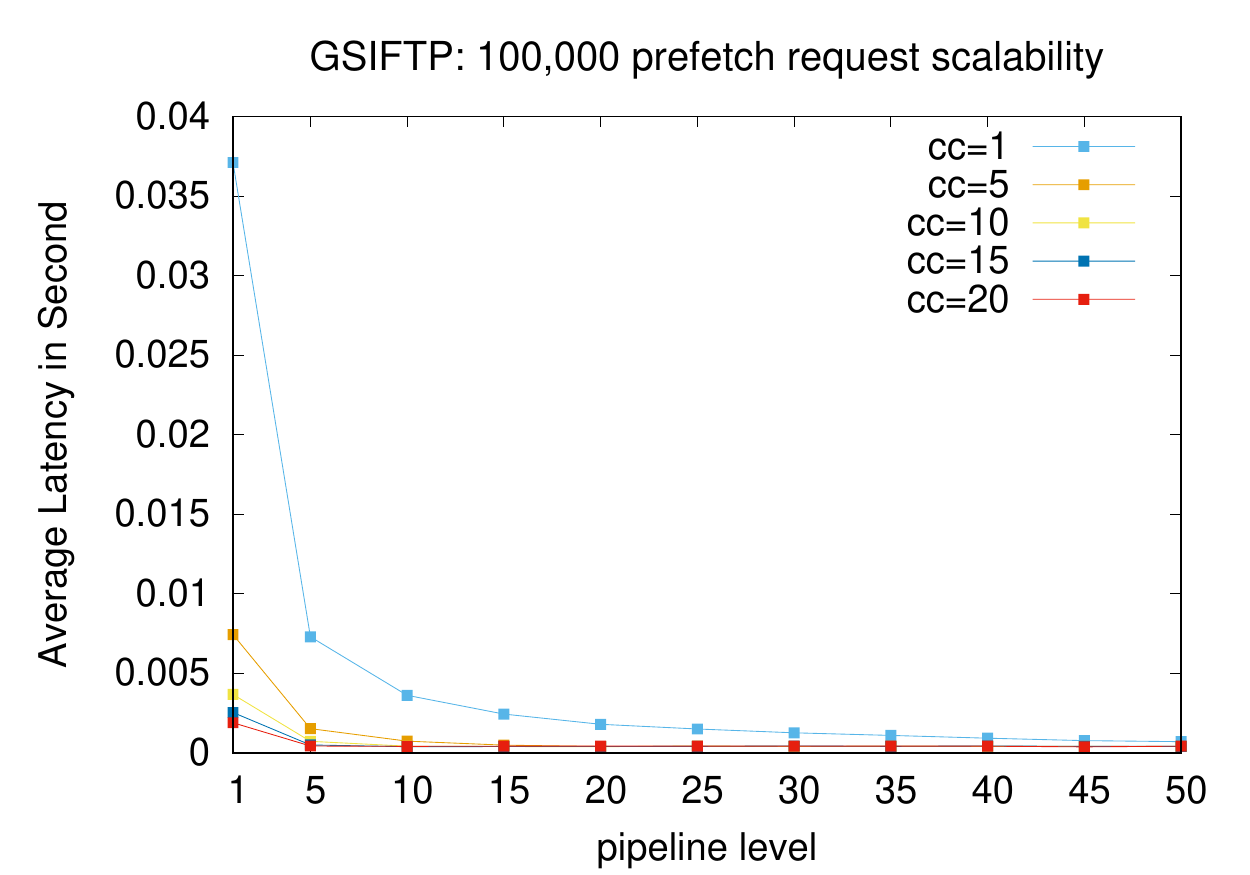} \\
    \includegraphics[keepaspectratio=true,angle=0,width=0.48\columnwidth]{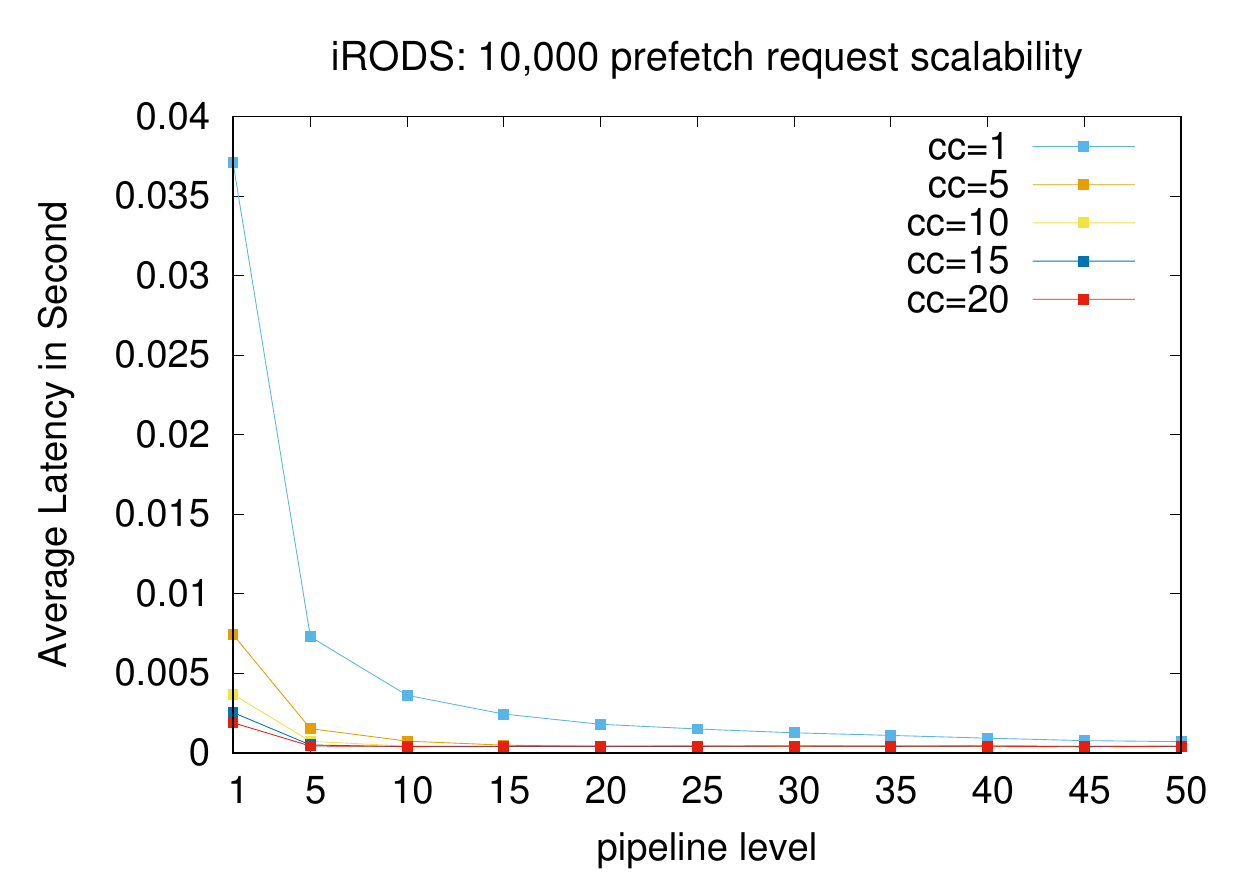}
 \includegraphics[keepaspectratio=true,angle=0,width=0.48\columnwidth]{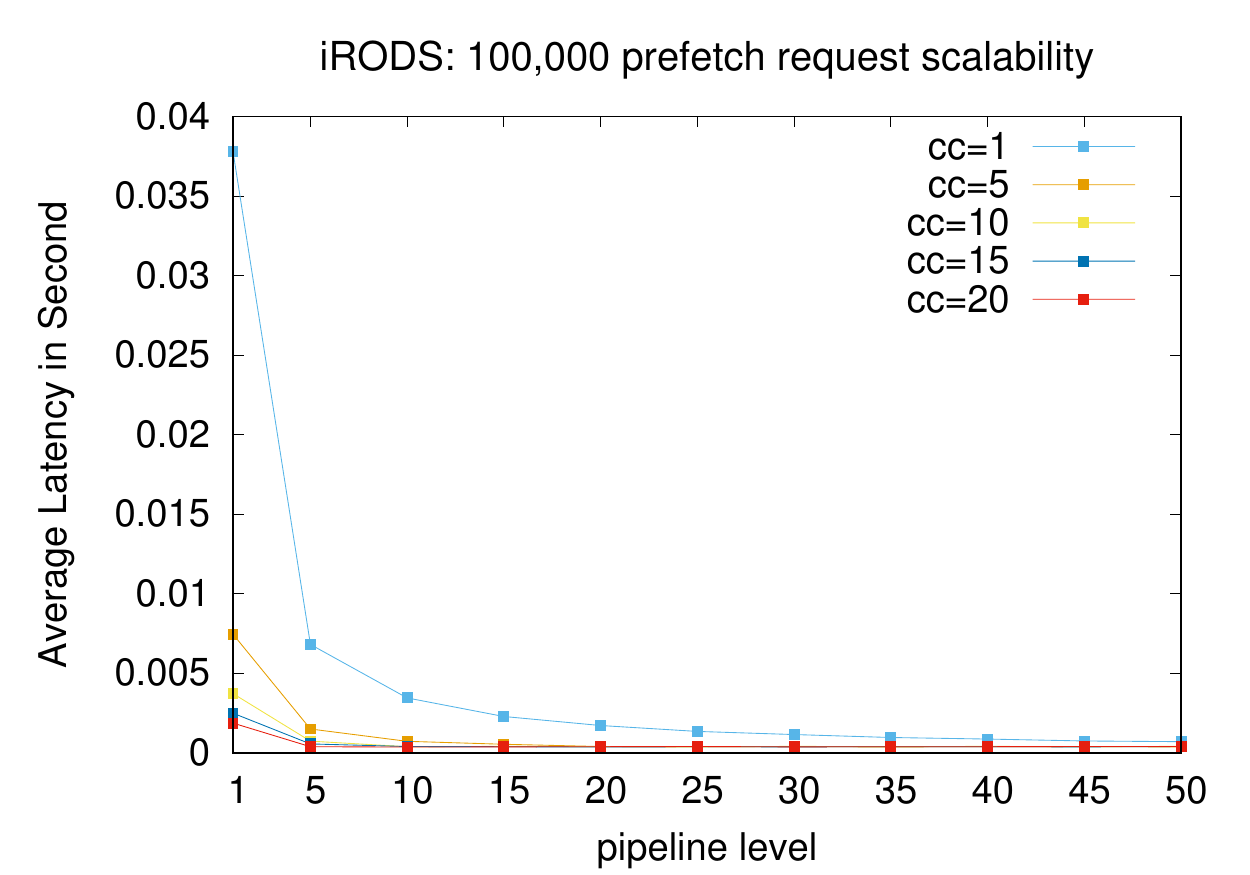} \\
    \includegraphics[keepaspectratio=true,angle=0,width=0.48\columnwidth]{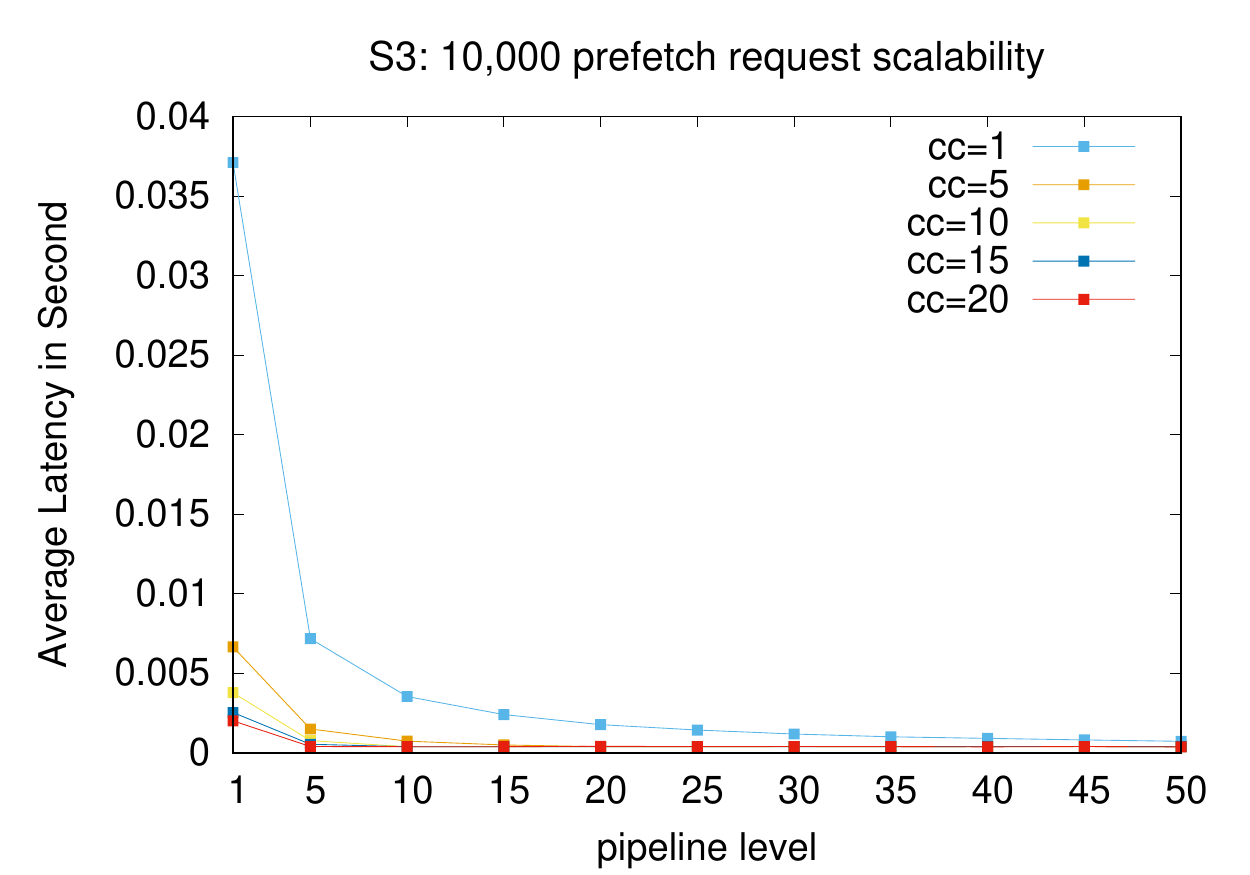}
 \includegraphics[keepaspectratio=true,angle=0,width=0.48\columnwidth]{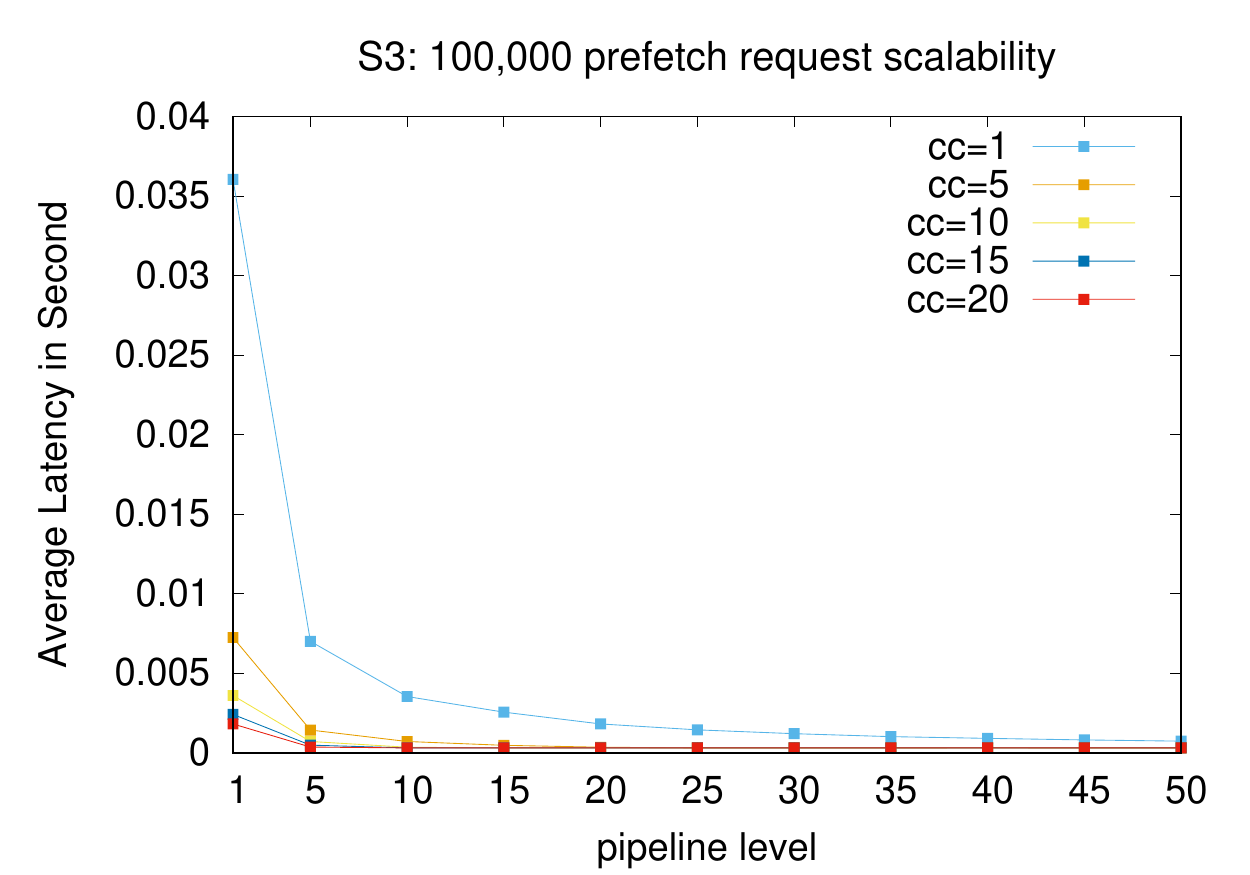} \\
\end{tabular}
\caption{Scalability of pipeline and concurrency.}
\label{fig:1w-pipeline}
\end{figure}

Figure~\ref{fig:1w-pipeline} demonstrates the scalability of prefetching files metadata from heterogeneous I/O servers. We turn off caching effects in the testbed and let one edge service initialize the sending of 10,000 and 100,000 distinct prefetch requests to the cloud, respectively, and calculated the average prefetching elapse time on the SMURF-Edge side. We continue to increase the number of concurrency channels and each channel's maximum pipeline capacity until there are no noticeable performance gains. The scalability performance between the different protocols is similar, and the SMURF system can reduce the prefetching latency to 0.6 millisecond per request on average, which means the system can complete the prefetching of 100,000 metadata contents in 60 seconds.

\begin{figure}[h]
\begin{tabular}{ccc}
 \includegraphics[keepaspectratio=true,angle=0,width=0.48\columnwidth]{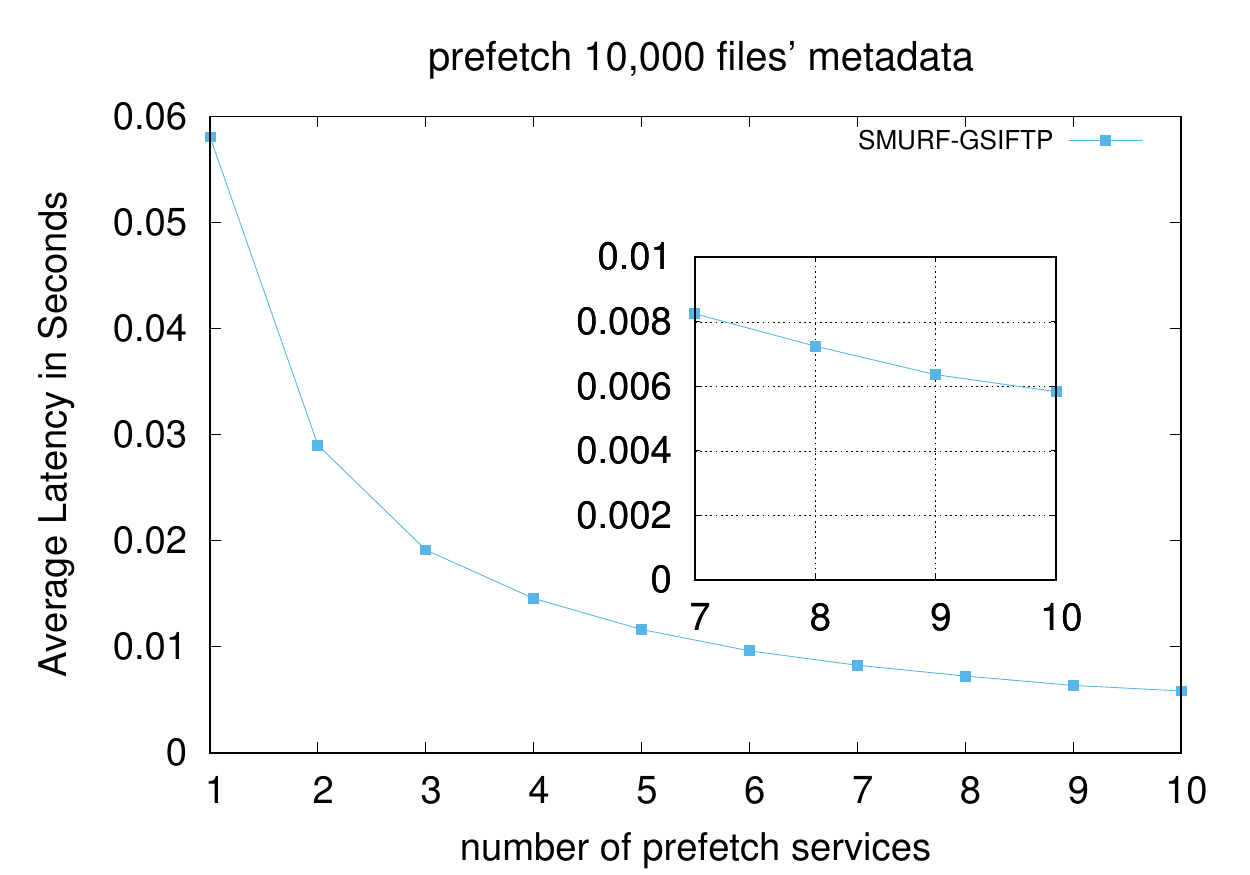}
 \includegraphics[keepaspectratio=true,angle=0,width=0.48\columnwidth]{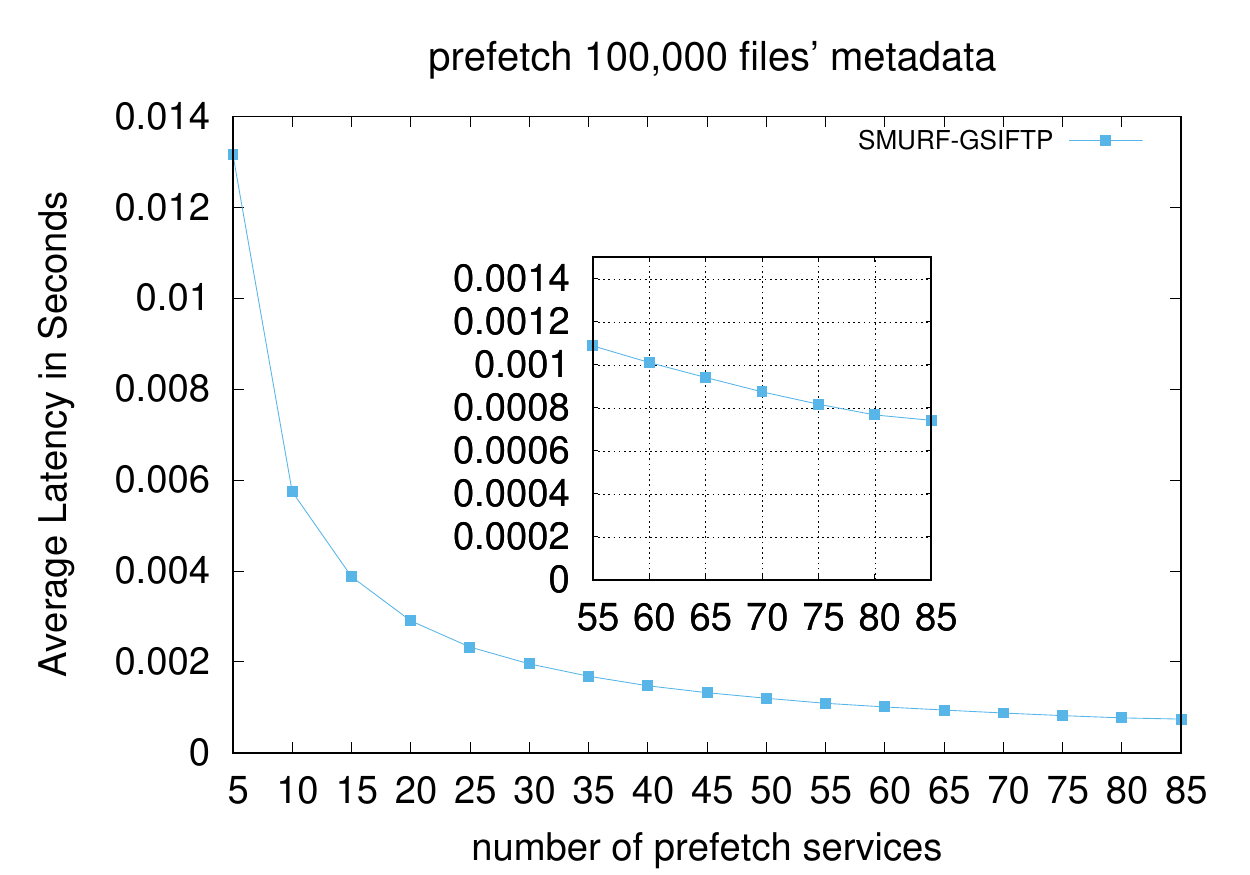}\\
\end{tabular}
\caption{Scalability of metadata prefetching from XSEDE-Comet@SDSC.}
\label {xsede-comet-gsiftp-prefetchscalability}
\vspace{-4mm}
\end{figure}

We also evaluated the scalability of prefetching from XSEDE Comet~\cite{xsede-comet} endpoints at San Diego Supercomputer Center (SDSC) in Figure~\ref{xsede-comet-gsiftp-prefetchscalability}, where the average prefetch latency per request is around 0.8 millisecond. Note that the transferring of files metadata over GSIFTP is conducted in the control channel by sending the command ``MLSC''. SMURF still supports FTP/GSIFTP data channel metadata transferring, which has been evaluated in our previous work~\cite{zhang2015dls}.

\subsubsection {Cache Hit Rate}

Figure~\ref{fig:yahoos3}(a) compares the cache hit rate between LRU cache and different prefetching schemes. Our directory locality scheme (denoted as ``DLS'') outperforms all other schemes on cache hit rate. It can achieve around 90\%+ on all individual Hadoop audit logs because the directory locality scheme can successfully capture the access pattern of the ``listStatus'' operation in Yahoo! Webscope Hadoop audit log. AMP is another prediction scheme with a high prediction rate that can achieve around 65\%+ accuracy. We train the AMP model on each day's trace and use that trained model for the next day's prefetching prediction. The AMP scheme's high prediction rate comes from the fact that there are many overlapping file paths of ``listStatus'' metadata operations between successive days. The first day (Part-00000) of AMP performance has been set to the same value as that of the LRU cache since no previous day data is available. In Yahoo! Webscope datasets, day four (Part-00003) data is not available; hence, we use the day three (Part-00002) trained model for conducting the AMP performance prediction on day four trace. We also found that the cache hit rates of Nexus and Farmer are almost the same as that of LRU cache (below 25\%) since the prefetching candidates suggested by the prediction schemes of Nexus and Farmer are all from the history requests. Simultaneously, the Hadoop audit log exhibits significantly skew popularity access in the ``listStatus'' metadata operation. Most of the ``listStatus'' operations execute on a file path once or occasionally, which inevitably causes the low prediction rate of prediction schemes based on history access sequence.

\subsection {Evaluation of Prefetch Schemes on Yahoo! Hadoop Grid Trace Logs}

\begin{figure}[t]
\includegraphics[keepaspectratio=true,angle=0,width=1\columnwidth]{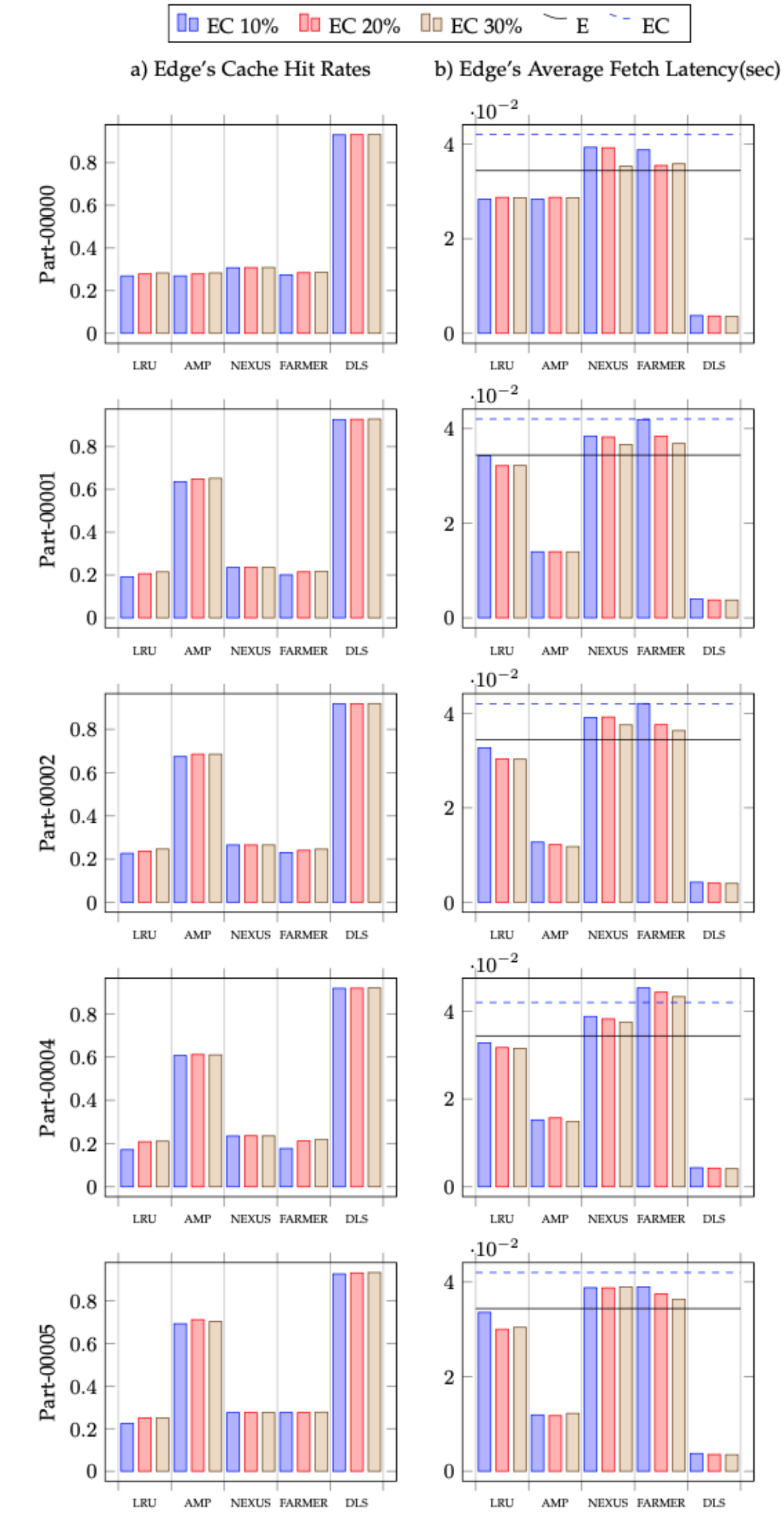}
\caption{Cache hit rate and average fetch latency between prediction schemes on Yahoo! Hadoop trace.}
\label{fig:yahoos3}
\vspace{-4mm}
\end{figure}

We conduct the experiments on the edge node and replay the trace logs with different settings. The experiments have been evaluated on the Edge-Cloud I/O path (the abbreviation term ``EC'' in Figure~\ref{fig:yahoos3}) with the prefetch schemes installation on the edge node. The cache size also has been taken into consideration, where the cache capacity on the edge node has been increased by the percentage (10\%, 20\%, and 30\%) of total requests in each trace log, and the average memory usages have been calculated by the oshi~\cite{oshi} software tool in Table~\ref{tab:prefetch-scheme-mem}. 
We also measure the average fetch time latency without the caching, and prefetching effects on the edge node denoted as E (Edge I/O path) and EC in Figure~\ref{fig:yahoos3}). In the following sections, we will compare and discuss the prefetch scheme's performance on the criteria of cache hit rate, average fetch latency, and storage usages, respectively.

\begin{table}[t]
\caption[Caption for LOF]{Prefetch Schemes Average Memory Usage (GB) on Edge Node.}
  \centering
  \begin{tabular}{|c|c|c|c|}
\hline
{\bf Prefetch Scheme} & {\bf 10\%} & {\bf 20\%}  & {\bf 30\%} \\
\hline
\hline
LRU & 13 & 19 & 22\\
\hline

AMP & 13 & 19 & 22\\
\hline

NEXUS & 15 & 21 & 25\\
\hline

FARMER & 15 & 21 & 25\\
\hline

DLS & 13 & 20 & 22\\
\hline
  \end{tabular}
  \label{tab:prefetch-scheme-mem}
\end{table}

\subsection {Performance of Continuum Caching}\label{continuum-cache}
Based on the evaluation of prefetch schemes, we decide to choose DLS as the default prefetch scheme and repeat the experiments on the Yahoo! Hadoop trace logs to evaluate the continuum caching performance on the Edge-Cloud and Edge-Fog-Cloud I/O paths. The edge node and the fog cluster are deployed in the same site, and the distributed cache system is installed across all the fog cluster nodes to provide the extra caching layer between the edge node and the cloud. We evaluate the average fetch latency and cache hit rate on the edge node between two aforementioned I/O paths with the increasing continuum caching capacity in Tables~\ref{tab:continuum-cache-averagefetchlatency} and~\ref{tab:continuum-cache-hitrate} and the average system memory usage in Table~\ref{tab:continuum-cache-mem}, where on the Edge-Cloud I/O path we first set the relatively small cache size at 0.5\% percentage of the total requests (around 20,000 metadata entries) on the edge node and keep increasing its cache size to 10\% until there are no obvious performance gains on the average fetch latency and cache hit rate. Accordingly, on the Edge-Fog-Cloud I/O path, we increase the fog cluster cache size to the percentages of 1\%, 5\%, and 10\%. It is clearly shown that the average fetch latency of the edge node is significantly reduced when the system sets a relatively larger cache size on the fog cluster. When the edge node has been configured with the constant cache size, namely, 0.5\% cache capacity, the fog cluster caching and prefetching can reduce the edge node average fetch latency up to 46\% and slightly increases the cache hit rate. This result comes from the fact that the nearby fog cluster can effectively cache and prefetch the demanded metadata shortly, and most cache-miss fetching requests on the edge node will be directly retrieved from the nearby fog cluster instead of the remote cloud. 

Compared with the EC settings, the average fetch elapse time has been delayed by the communication overhead between the edge node and the fog cluster but can be reduced by increasing the caching capacity on the fog cluster. When the fog cluster has been configured with 10\% cache capacity, the average fetch latency of the edge node with 0.5\% cache capacity can be almost the same as that of the edge node with 10\% cache capacity.

\subsubsection{Average Fetch Latency}

In Figure~\ref{fig:yahoos3}(b), we calculate the average fetch time between the prefetch schemes and measure the accumulated overhead of metadata transferring without any cache and prefetch installed. The setting of ``E'' (denoted with the solid horizontal line) shows the average latency of fetching performance that the edge node directly fetches metadata from the remote I/O server. The average fetching latency of the ``EC'' setting is denoted with the dashed horizontal line. The performance of LRU and all prefetch schemes are usually below the ``EC'' bar, which demonstrates that caching and prefetching is still an effective way to reduce the average fetching latency even with the low cache hit rates.

The prefetching scheme with a higher prediction rate can significantly reduce the average fetch latency since most metadata can be accessed locally. With the highest prediction rate (90+\%) of the DLS prefetch scheme, the average fetching latency can be reduced to 0.004 seconds. The AMP (65+\%) can achieve the average fetching latency of 0.015 seconds. Note that we have to use external storage to store the AMP training model, but the AMP model's high prediction rate can offset the overhead of database operations. The LRU with a larger amount of cache size can slightly reduce the average fetching latency. 
Nexus and Farmer's average latencies are relatively higher (above the solid bar). This may be due to two factors: 1) The RTT between client and remote I/O server is around 32 milliseconds, while the accumulated RTT of EC path is above 40 milliseconds; 2) Nexus and Farmer prediction rates are nearly the same as that of LRU cache, but there is extra computation to build relation graph on the fly. Moreover, the overhead of constructing and updating the relation graph in the Nexus and Farmer prefetch schemes is not ignorable.

\begin{table*}[htbp]
\caption{Edge Node Average Fetch Latency (milliseconds) Scalability with Increasing Continuum Caching Capacity.}
  \centering
  \begin{tabular}{|c|c|c|c|c|c|c|c|}
\hline
{\bf Log Name} & {\bf EC 0.5\%} & {\bf EC 1\%}  & {\bf EC 5\%} & {\bf EC 10\%} & {\bf E 0.5\% F 1\%} & {\bf E 0.5\% F 5\%} & {\bf E 0.5\% F 10\%}\\
\hline
\hline
part-00000 & 5.9 & 4.7 & 3.8 & 3.8 & 7.0 & 4.4 & 3.8\\
\hline

part-00001 & 6.4 & 4.8 & 4.0 & 4.0 & 6.2 & 4.5 & 4.4\\
\hline

part-00002 & 4.7 & 4.4 & 4.3 & 4.2 & 4.5 &  4.3 & 4.2\\
\hline

part-00004 & 5.4 & 5.0 & 4.4 & 4.3 &  5.2 & 5.0 & 4.7\\
\hline

part-00005 & 5.7 & 5.3 & 4.1 & 3.7 &  7.8 & 5.5 & 4.3\\
\hline
  \end{tabular}
  \label{tab:continuum-cache-averagefetchlatency}
\end{table*}

\begin{table*}[htbp]
\caption{Edge Node Cache Hit Rate (percentage) with Increasing Continuum Caching Capacity.}
  \centering
  \begin{tabular}{|c|c|c|c|c|c|c|c|}
\hline
{\bf Log Name} & {\bf EC 0.5\%} & {\bf EC 1\%}  & {\bf EC 5\%} & {\bf EC 10\%} & {\bf E 0.5\% F 1\%} & {\bf E 0.5\% F 5\%} & {\bf E 0.5\% F 10\%}\\
\hline
\hline
part-00000 & 83\% & 88\% & 93\% & 93\% & 77\% & 82\% & 84\% \\
\hline

part-00001 & 85\% & 88\% & 93\% & 93\% & 79\% & 87\% & 88\%\\
\hline

part-00002 & 89\% & 90\% & 92\% & 92\% & 88\% & 89\% & 89\%\\
\hline

part-00004 & 84\% & 88\% & 92\% & 92\% &  83\% & 83\% & 84\% \\
\hline

part-00005 & 78\% & 81\% & 88\% & 93\% & 73\% & 78\% & 77\% \\
\hline
  \end{tabular}
  \label{tab:continuum-cache-hitrate}
\end{table*}

\section{Related Work }



Metadata prefetch prediction~\cite{li2012real, gu2006nexus, 4534250} studies developed different strategies to predict future requests as accurately as possible. NEXUS~\cite{gu2006nexus} applies a weighted-group-based prefetching algorithm to prefetch prediction. A weighted directed graph is built on the fly when the metadata server (MDS) receives requests from clients. The proposed polynomial time complexity algorithm looks up and analyzes requests in a predefined capacity history window. For a given access sequence, the history queue is populated with each request in the access sequence order. Each enqueued request is created as a vertex in this weighted relationship graph,  where a directed edge from any queueing request connects to this newly enqueued request. The weight of each edge connection is calculated according to the successor relationship strength. For a given request, the prediction predictor looks up the graph to find out the directly connected vertices (requests) and predict top-k vertices with the largest edge weight as the best prefetching candidates. Experiments show that their prefetching prediction can effectively reduce clients' average response time with reasonable overhead.

FARMER~\cite{xia2008farmer} investigates how a request's attributes information (e.g., ``Host'', ``UserID'', ``ProcessID'' and file path ) can affect the file successor probability (the likelihood of successor being accessed if the predecessor has been accessed). The authors statistically analyze the average probabilities for the different trace sequences. They conclude that the same attribute will have a different successor probability between various traces. The access pattern without considering the semantic attributes is not sufficient to predict the file access probability. They apply a linear combination model to consider the combined effect of the history access sequence and the semantic attributes of requests. FARMER builds a relationship graph between predecessor and successors in a specific size history window similar to NEXUS and applies Integrated Path Algorithm (IPA) to detect the semantic attributes correlation between predecessor and each successor. The best prefetching accuracy that FARMER achieves is 64\%, where NEXUS can perform 43\%.

AMP~\cite{4534250} uses a different approach to predict the request pattern based on the analysis of the historical access sequence. The authors apply N-gram~\cite{brown1992class} model, which has been widely used in natural language processing, to train the prediction model in a quasi-online fashion (overnight training and use training result for the next day's prefetching prediction). AMP states that a 3-gram model can have more constraints on prediction and give more accurate predictions. They also claim that a 3-gram with up to 6 prefetching items can achieve a better hit ratio with less computation overhead. The experiments show that AMP can outperform NEXUS by 4\% on hit ratio and reduce the average response time by 8\%.


Hierarchical caching has been well studied in the literature. For the web caching systems, Wolman et al. ~\cite{wolman1999scale} conducted their analysis on the hierarchical tree structure cache and evaluated the advantages and drawbacks of inter-proxy cooperation to demonstrate the performance benefits of cooperative caching. Sadeghi et al. ~\cite{sadeghi2019deep} represented the popular tree hierarchical cache networks into a two-level network caching, where the network of caching nodes has been managed in a two-timescale approach. The researchers formulated the file transmission cost model as the Markov decision process (MDP) and propose a novel reinforcement learning (RL) to select the efficient caching policy to adapt to the dynamic evolution of file requests and caching policies of the network nodes. Jia et al. ~\cite{jia2005optimal} considered a cached content placement problem in
a hierarchical web proxies environment. The authors formulated the problem to minimize the data access costs by considering the distance between the source of requests and the closest destination with the requested data. Tran et al~\cite{tran2017cooperative} introduced a novel cooperative hierarchical caching framework under the C-RAN~\cite{chih2014recent} architecture. Inside the proposed framework, the complementary cloud cache and edge caches have been managed by a centralized controller at the cloud. The authors evaluated the performance by configuring the cache installation on the different hierarchical layers. The experiments show that the proposed framework significantly outperforms traditional edge-only caching schemes.

\section {Conclusion}

This paper addresses two crucial IoT research challenges in accessing remote metadata: heterogeneity and scalability. We have presented a novel solution for efficient and scalable metadata access for distributed and heterogeneous applications across wide-area networks, called SMURF. Our solution combines novel pipelining and concurrent transfer mechanisms with reliability, provides distributed continuum caching and prefetching strategies to sidestep fetching latency, and achieves scalable and high-performance metadata fetch/prefetch services in the cloud. We also studied the applicability of semantic locality in real trace logs, which is not well utilized in traditional metadata access prediction techniques, and implemented a novel prefetch predictor based on semantic locality. We compared it with three existing state-of-the-art prefetch schemes (NEXUS, FARMER, and AMP) on Yahoo! Hadoop audit traces. Our experimental results show that SMURF can achieve 90\% accuracy during prefetch prediction and reduce the average fetch latency up to 50\% compared to the other mechanisms.


\bibliographystyle{IEEEtran}
\bibliography{IEEEabrv,mybibfile}

\end{document}